\makeatletter
\declare@file@substitution{revtex4-1.cls}{revtex4-2.cls}
\makeatother
\documentclass[aps,prd,twocolumn,showpacs,superscriptaddress,groupedaddress,nofootinbib]{revtex4-2}
\usepackage[portuges, english]{babel}
\usepackage[utf8]{inputenc}
\usepackage{amsmath}
\usepackage[makeroom]{cancel}
\usepackage{slashed}
\usepackage{enumitem}
\usepackage{graphicx}
\usepackage[caption=false]{subfig}
\usepackage{natbib}
\usepackage[hidelinks,colorlinks=true,linkcolor=red,citecolor=red]{hyperref}

\usepackage{tikz,hyperref, xcolor}
\definecolor{lime}{HTML}{A6CE39}
\definecolor{Emerald}{HTML}{50c878}
\definecolor{PineGreen}{HTML}{01796F}
\definecolor{ForestGreen}{HTML}{228B22}
\definecolor{Coral}{HTML}{FF7F50}
\definecolor{YellowOrange}{HTML}{E94E16}
\hypersetup{linkcolor=Emerald,citecolor=red,filecolor=red,
urlcolor=blue}

\def\adv{AIP Advances}%

\begin{document}

\medskip\noindent

\title{\Large  Bosonic Dark Matter Dynamics in Hybrid Neutron Stars}

 \DeclareRobustCommand{\orcidicon}{%
 	\begin{tikzpicture}
 		\draw[lime, fill=lime] (0,0) 
 		circle [radius=0.16] 
 		node[white] {{\fontfamily{qag}\selectfont \tiny ID}};
 		\draw[white, fill=white] (-0.0625,0.095) 
 		circle [radius=0.007];
 	\end{tikzpicture}
 	\hspace{-2mm}
 }
 \foreach \x in {A, ..., Z}{%
 	\expandafter\xdef\csname orcid\x\endcsname{\noexpand\href{https://orcid.org/\csname orcidauthor\x\endcsname}{\noexpand\orcidicon}}
 }
\def\adv{AIP Advances}%
\newcommand{\orcidauthorB}{0000-0003-4777-4188} 
\author{Zakary Buras-Stubbs\orcidB{}}
\affiliation{Centro de Astrof\'{\i}sica e Gravita\c c\~ao  - CENTRA, \\
	Departamento de F\'{\i}sica, Instituto Superior T\'ecnico - IST,\\
	Universidade de Lisboa - UL, Av. Rovisco Pais 1, 1049-001 Lisboa, Portugal}
\email{zburasstubbs@tecnico.ulisboa.pt}
%
\newcommand{\orcidauthorA}{0000-0002-5011-9195} 
\author{Ilídio Lopes\orcidA{}}
\affiliation{Centro de Astrof\'{\i}sica e Gravita\c c\~ao  - CENTRA, \\
	Departamento de F\'{\i}sica, Instituto Superior T\'ecnico - IST,\\
	Universidade de Lisboa - UL, Av. Rovisco Pais 1, 1049-001 Lisboa, Portugal}
\email{ilidio.lopes@tecnico.ulisboa.pt}




\begin{abstract}
\noindent	
This research studies the intricate interplay between dark and baryonic matter within hybrid neutron stars enriched by anisotropic bosonic dark matter halos. Our modelling, guided by the equation of state with  a free parameter, reveals diverse mass-radius correlations for these astronomical objects. A pivotal result is the influence of dark matter characteristics—whether condensed or dispersed—on the observable attributes of neutron stars based on their masses. Our investigation into anisotropic models, which offer a notably authentic representation of dark matter anisotropy, reveals a unique low-density core halo profile, distinguishing it from alternative approaches. Insights gleaned from galactic clusters have further refined our understanding of the bosonic dark matter paradigm.
Observational constraints derived from the dynamics of galaxy clusters 
have been fundamental in defining the dark matter particle mass to lie between 0.05 GeV and 0.5 GeV and the scattering length to range from 0.9 fm to 3 fm.
Using terrestrial Bose-Einstein condensate experiments, we have narrowed down the properties of bosonic dark matter, especially in the often overlooked 3 to 30 GeV mass range. Our findings fortify the understanding of dark and baryonic matter synergies in hybrid neutron stars, establishing a robust foundation for future astrophysical pursuits.
\end{abstract}


\maketitle

\section{Introduction} \label{sec:intro}

\medskip\noindent
Dark matter, one of the most intriguing and mysterious phenomena in astrophysics, remains a subject of ongoing investigation \citep[e.g.,][]{2017arXiv170704591B,2022RPPh...85e6201B}.  This invisible form of matter, which does not emit, absorb, or reflect electromagnetic radiation, accounts for roughly 27\% of the Universe's total mass-energy content \citep[e.g.,][]{2014IJMPD..2330005D}. The presence of dark matter has been deduced from its gravitational influence on galaxy movements, the vast cosmic structures, and the cosmic microwave background radiation 
\citep[e.g.,][]{2017FrPhy..12l1201Y}.
 
 \medskip\noindent
It is widely accepted that dark matter cannot be composed of fundamental particles within the Standard Model of particle physics, and it is more likely that dark matter is made up of particles that have not yet been discovered.
 Various hypothetical particle candidates have been proposed as potential solutions to the dark matter problem, and these particles can generally be classified into several categories based on their mass \citep{2010ARA&A..48..495F,2015PhR...555....1B,2018RPPh...81f6201R,2019Univ....5..213P}. These categories include fuzzy dark matter or axion-like particles, axions, sterile neutrinos, and weakly interacting massive particles (WIMPs).

 \medskip\noindent
Our focus in this work is to investigate how the structure of neutron stars can be affected by dark matter particles within the WIMPs and axion mass ranges. In particular, we will be focusing on particles in the $10^{-2}$ --  $10^2$  GeV range.

\medskip\noindent
There are several papers in the literature that discuss the relationship between dark matter and neutron stars. Some recent examples include \citep{2008PhRvD..77b3006K,2010PhRvD..82f3531K}, which focus on general self-annihilating WIMP models; \citep{2017PhRvD..96h3004P,2020PhRvD.102f3028I,PhysRevC.89.025803, PhysRevD.99.043016, 2020JPhG...47i5202Q, PhysRevD.104.063028, PhysRevD.104.123006, PhysRevD.105.023008, PhysRevD.105.043013, PhysRevD.106.043010, 2022PhRvD.105l3034D}, which focus on specific fermionic dark matter models; \citep{2018IJMPD..2750093P,PhysRevD.105.023001}, that consider Bose-Einstein Condensate (BEC) dark matter; and \citep{PhysRevD.97.123007} where both fermionic and BEC models are studied. These papers explore the possible effects of dark matter on neutron star properties and the potential for dark matter detection through observations of neutron stars. Additionally, there are several reviews that provide a more comprehensive overview of the field, such as \citet{2021ARNPS..71..433L} and \citet{2020Univ....6..222D}.

\medskip\noindent
Anisotropic matter exhibits different properties or behaviours depending on the direction
\citep[for a review see][]{1997PhR...286...53H,2022NewAR..9501662K}.
 Such behaviour can lead to pressure differences in various directions within neutron stars or other astronomical objects \citep[e.g.,][]{2016arXiv161102253G,2021JPhCS1816a2025R,2021PhRvD.103h4023M}. This characteristic influences how matter interacts with surrounding objects, affecting their formation, structure, and evolution, shaping our understanding of dark matter. For instance, dark matter clouds are expected to exhibit local anisotropy, similar to any collisionless system of particles. Researchers have extensively studied these systems, especially in the context of galaxy dynamics
 \citep[e.g.,][]{1963MNRAS.125..127M,1987gady.book.....B,1991MNRAS.253..414C}.

\medskip\noindent
The theory of anisotropic fluids in General Relativity is well-established. Past research has demonstrated that anisotropic fluids could be geodesic in general relativity \citep{2002JMP....43.4889H}. A comprehensive study of spherically symmetric dissipative anisotropic fluids has been presented, and exact static spherically symmetric anisotropic solutions of field equations have been obtained and analyzed by \citet{2004PhRvD..69h4026H}, 
\citet{1982PhRvD..26.1262B}, and \citet{2008PhRvD..77b7502H}. Furthermore, calculations of anisotropic stars in general relativity and their mass-radius relations have been conducted \citet{2003RSPSA.459..393M}. A detailed collection of recent articles on this topic is available on 
 \citet{2022NewAR..9501662K}.

\medskip\noindent
This study introduces a novel astrophysical object: a hybrid compact star composed of a neutron star within an anisotropic dark matter halo. We will investigate the unique characteristics of these models to gain a deeper understanding of their formation. By examining the properties of dark matter, we propose that anisotropic dark matter may play a vital role in forming these hybrid compact stars. Our analysis focuses on a static, spherically symmetric configuration featuring a neutron star enveloped by dark matter. Analogous to the baryonic matter within the neutron star, the properties and distribution of the dark matter, positioned at a specific distance from the star's centre, are described using the Tolman-Oppenheimer-Volkoff (TOV) equations.

\medskip\noindent
We assume that the dark matter envelope does not interact with the neutron star's baryonic matter within the envelope. We model how dark matter affects neutron star observations using the theory of compact stars. We focus on two main assumptions: first, dark matter with a nonzero radial pressure term in the TOV equations, $P_{r\chi}(r) = P_\chi(r)$, and second, an anisotropic energy-momentum tensor leading to a nonzero tangential pressure, $P_{\perp\chi}(r)$. We justify these choices and describe the dark matter distribution by calculating $\Pi_\chi (r)=P_{\perp\chi}(r) - P_{r\chi}(r)$.

\medskip\noindent
In this study, we concentrate on anisotropic dark matter's effects on neutron stars within the dark matter halo, examining the mass-radius relation. Furthermore, we assume that the dark matter particles are bosons. 
We compare the dark matter density profiles and mass-radius relations to those with isotropic dark matter and a neutron star in a vacuum.

 \medskip\noindent 
In this article, we structure our presentation as follows. We begin with a summary of current research topics on neutron stars and dark matter in this section. In Section \ref{sec:PNSADM}, we provide an overview of the hybrid model employed in this study. Subsequently, in Section \ref{sec:EOSADM}, we discuss the properties of the equation of state for neutron stars and anisotropic dark matter. In Section \ref{sec:DNR}, we investigate the properties of dark matter halos formed around neutron stars. In Section \ref{sec:PDMP}, we examine constraints on the properties of dark matter particles. Finally, we summarize our findings and present our conclusions in Section \ref{sec-Con}.

 \medskip\noindent   
 Here, unless specified otherwise, we utilize natural units, where the speed of light ($c$), the Boltzmann constant ($k_{B})$ and the reduced Planck constant ($\hbar$) are set to $1$. Through standard conversion rules, all conventional units can be expressed in terms of GeV. The most frequently used units in this work are:
 1 centimetre = $ 5.05 \times 10^{13}\; {\rm GeV^{-1}}$, 
 1 gram = $ 5.610  \times 10^{23}\; {\rm GeV}$ and 	
 1 second = 	$	1.519 \times 10^{24}\; {\rm GeV^{-1}}$.

\section{Properties of Neutron Stars and Anisotropic  Dark matter}  
\label{sec:PNSADM}  

 \medskip\noindent 
Here we start by briefly summarising the equations governing the components of our admixed neutron star.

\subsection{Tolman–Oppenheimer–Volkoff Equations}

 \medskip\noindent 
We identify that the action governing our system will essentially take the form
\begin{equation} \label{eq: S}
	S = S_G+S_M,
\end{equation}
for which $S_G$ defines the gravitational component, and hence is given by the usual Einstein-Hilbert  action, and $S_M$ encapsulates matter terms associated with a perfect fluid whose properties are fully determined by its energy density $\rho$, its radial pressure $P_r$, its tangential pressure $P_\perp$, as well as an equation of state defined such that $F(\rho,P_r) = 0$ 
\citep[e.g.,][]{2019EPJP..134..454L}.

\medskip\noindent
From this derivation, by varying with respect to the metric, we obtain the standard Einstein field equations without the inclusion of a cosmological constant:
\begin{equation}
\label{eq: EEq}
G_{\mu\nu} = R_{\mu\nu} - \frac{1}{2}g_{\mu\nu}R = 8 \pi T_{\mu\nu}
\end{equation}
Here, Newton's constant G is set equal to unity. 
The total stress-energy tensor can be expressed as:
\begin{equation} \label{eq: Tmn}
	T^\mu_\nu = {\rm Diag}\left(-\rho, P_r, P_\perp, P_\perp\right).
\end{equation}
In the comoving frame, the physical matter content consists of an anisotropic fluid with energy density $\rho$, radial pressure $P_r$, and tangential pressure $P_\perp$.

\medskip\noindent
We define a general metric tensor as follows:
\begin{equation} \label{eq:g}
ds^2= -e^\nu dt^2 + e^\lambda dr^2 + r^2d\Omega^2,
\end{equation}
where $\nu (r)$ and $\lambda(r)$ are the metric potentials, depending only on the radial coordinate, and $d\Omega^2 \equiv (d\theta^2 + \sin^2{\theta}\;d\phi^2)$ represents the solid angle element \citep[e.g.,][]{2007MNRAS.375.1265S}. By solving Einstein's equations, we find that:
\begin{equation} \label{eq: B}
	e^{\lambda(r)} = \frac{1}{1 - \frac{2m}{r}},
\end{equation}
and
\begin{equation} \label{eq: lam}
	\nu^\prime(r) = \frac{2 m + 8 \pi r^3 P_r}{r^2 - 2 m r }.
\end{equation}
Moreover, we obtain the hydrostatic equilibrium equation \citep{1939PhRv...55..374O,1939PhRv...55..364T}, also referred to as the generalized Tolman–Oppenheimer–Volkoff (TOV) equation:
\begin{equation} \label{eq:TOV}
	\frac{dP_{r}}{dr} = -\frac{(\rho + P_{r})(m + 4\pi r^3 P_r)}{r(r - 2m)} + \frac{2\Pi}{r},
\end{equation}
where $\Pi \equiv P_{\perp} - P_{r}$. It is important to note that for an isotropic substance, we have $\Pi = 0$ or $P_{\perp } = P_{r}$. 
In our calculations for the baryonic component, we will enforce this condition, causing the final term to vanish and providing the hydrostatic equilibrium equation, also known as the generalised TOV equation. 
\\Within this a natural interpretation of the calculations we have performed becomes manifest. Recognising that hydrostatic forces are described by $F_h = -\frac{dP_{r}}{dr}$, gravitational forces by $F_g = -\frac{(\rho + P_{r})(m + 4\pi r^3 P_r)}{r(r - 2m)}$, and anisotropic forces are given by $F_a = \frac{2\Pi}{r}$, then it becomes clear that equation \eqref{eq:TOV} is just the statement that
\begin{equation}\label{eq: fsum}
	F_g + F_h + F_a = 0.
\end{equation}
\\The total mass, $m(r)$, is computed using the standard differential equation relating mass and density:
\begin{equation} \label{eq: dmdr}
	\frac{dm}{dr} = 4\pi r^2 \rho.
\end{equation}
\\In accordance with a prevalent approach in the literature, as seen in the works of  \citet{2017PhRvD..96b3002P, 2011PhLB..695...19C, 2009APh....32..278S}, we assume that the interaction between dark matter and the baryonic matter is so weak that it can be reasonably neglected. Therefore, we consider the fluid to have two components: a baryonic component ($B$) representing neutron star matter and a non-baryonic dark matter ($\chi$) component. For instance, we define $\rho = \rho_B + \rho_\chi$, where $\rho_B$ is the energy density associated with standard baryonic matter ($B$) and $\rho_\chi$ is related to dark matter ($\chi$). Consequently, we assume that baryonic matter ($B$) and dark matter ($\chi$) are coupled solely through gravity, with their energy-momentum tensors conserved separately. As a result, the previous system of equations \eqref{eq:TOV} and \eqref{eq: dmdr} is split into two components:	
\begin{equation} \label{eq:TOVi}
	\frac{dP_{ri}}{dr} = -\frac{(\rho_i + P_{ri})(m + 4\pi r^3 P_{r})}{r(r - 2m)} + \frac{2\Pi_{i}}{r},
\end{equation}
where $\Pi_i=P_{ri}-P_{\perp i}$, and $P_r = P_{rB}+P_{r\chi}$,
accompanied by the additional equation:
\begin{equation} \label{eq: dmdri}
	\frac{dm_i}{dr} = 4\pi r^2 \rho_i.
\end{equation}
We see that our model reproduces the TOV equations found in the literature, \citep[e.g.,][]{1975A&A....38...51H,2020PhRvD.102f3028I}, where the effects of anisotropies only appears as an additive term, as one would intuit from equation \eqref{eq: fsum}.
\\
\\
In the aforementioned equations, the subscript index $i=B$ and $i=\chi$ denote baryonic matter ($B$) and dark matter ($\chi$), respectively. Therefore, $\rho_i$, $P_{ri}$, and $P_{\perp i}$ correspond to the density and pressure terms of the $i-$th component, where $r$ represents a purely radial component, and $\perp$ a tangential one. However, in our study, we assume that the baryonic matter is isotropic, meaning that $P_{\perp B}(r)=P_{rB}(r)$ and, as a result, $\Pi_B (r)=0$. The gravitational mass, $m$, is the sum of the masses of both components, that is, $m(r)=m_B(r)+m_\chi(r)$. The system of equations \eqref{eq:TOVi}, along with equation \eqref{eq: dmdri}, represents a generalization of equations \eqref{eq:TOV} and \eqref{eq: dmdr}.
\\We determine the radial density profiles of baryonic matter ($B$) and dark matter ($\chi$) using Equations \eqref{eq:TOVi} and \eqref{eq: dmdri} by applying two central and two boundary conditions for the star. The first set of conditions relates to the central densities of $B$ and $\chi$, while the second set ensures hydrostatic equilibrium for each matter component at their respective boundaries, forming spheres with radii $R_B$ and $R_\chi$ containing all the $B$ and $\chi$ mass components, respectively. Generally, $R_B \neq R_\chi$. Hydrostatic equilibrium is independently maintained for  $B$ and $\chi$  by the conditions:
\begin{equation}
	\label{eq:pR}
	P_i(R_i)=0.
\end{equation}
The $\chi$ component cannot be directly observed, so it is reasonable to identify the neutron star radius $R$ with $R_B$, i.e., $R_B=R$, where $R$ represents the observed radius of the neutron star. We choose to define our radius like this so that our analysis may be compared with observational data, as the neutron stars will have a visible radius in the usual sense, with an additional hidden mass. The total gravitational mass and the fraction of $\chi$ inside the neutron star are defined as:

\begin{equation}
	\label{eq:Mt}
	M_T=M_B(R_B)+M_\chi(R_\chi)
\end{equation}
and
\begin{equation}
	\label{eq:fchi}
	f_\chi=\frac{M_\chi(R_\chi)}{M_T}.
\end{equation}
It is evident that by varying the central densities of baryonic matter ($B$) and dark matter ($\chi$), we can obtain different values of the total gravitational mass $M_T$ and neutron star radius $R$ for a given $\chi$ fraction $f_\chi$.

\section{Equations of State and Anisotropic Dark Matter}  
\label{sec:EOSADM}

\subsection{Microphysics Within Neutron Stars}
 In our study of hybrid stars, we aim to accurately depict the complex physics inside neutron stars at various densities and temperatures. Our chosen microscopic model is based on a comprehensive framework suitable for the entire high-energy Quantum Chromodynamics (QCD) range. The model covers various temperatures and densities, considering different levels of isospin and strangeness. Notably, this model is consistent with known requirements for compact stars \cite[e.g.,][]{2019NuPhA.982..883S,2021PhRvL.126f1101A,2021ApJ...918L..29R} and findings from collider experiments \cite[e.g.,][]{2015PhRvL.114t2301P}, like those based on Lattice QCD. We mainly focus on the requirements of cold, dense stars, the behaviour of evenly distributed nuclear matter, and the conditions of high-temperature QCD at very low or no density. Moreover, this model also clearly explains how chiral symmetry is restored and how quarks are deconfined, supporting data from lattice QCD \cite[e.g.,][]{2017PhRvD..95e4504B}, perturbative QCD \cite[e.g.,][]{2014JHEP...05..027H} , and high-energy collider tests.

\medskip\noindent
Addressing all aspects above, we have chosen an equation of state computed within the framework of the Chiral Mean Field (CMF) model.  This model is grounded in the three-flavour chiral Lagrangian for hadronic matter expanded to encompass neutron stars \cite{Papazoglou:1998vr,Dexheimer:2008ax}.
In our study, we utilise a version of the CMF model augmented by Motornenko {\it et. al.}, as detailed in their article \cite{PhysRevC.101.034904,Motornenko:2020yme}.
This enhanced model encompasses a broad spectrum of QCD degrees of freedom.   The model includes, beyond protons and neutrons, hyperons and their parity partners, as well as a comprehensive array of hadronic resonances (encompassing both strange and non-strange baryons and mesons), as catalogued in the Particle Data Book  \cite[e.g.,][]{PTEP2022}. Additionally,  this model  incorporate the thermal effects of deconfined quarks and gluons as per the PNJL model \cite{ParticleDataGroup:2020ssz}. 
 When combined with electrons,  this represents the most comprehensive range of QCD degrees of freedom for the high-density equation of state.  Furthermore, the model enables the depiction of nuclear matter in a low-density regime, specifically for densities below  $10^{-2}~ n_{\rm sat}$,  which are prevalent in binary neutron star mergers \cite{Schneider:2017tfi}.
 
\medskip\noindent 
 This comprehensive approach enables the calculation of a unique equation of state for nuclear matter valid simultaneously in heavy ion collisions and binary neutron star mergers without the need for additional adjustments that might arise from using different and potentially inconsistent equations of state for distinct nuclear matter scenarios.

\medskip\noindent
The equation of state version used in our study effectively simulates a crossover transition for deconfinement at finite and zero density,  as indicated by lattice QCD~\cite{Motornenko:2018hjw,Motornenko:2020yme}. It accurately represents hadrons in medium, nuclei, nuclear matter, and neutron stars. This model generates a realistic nuclear ground state with these properties: saturation at a baryon number density of ${\rm n_{\rm sat}} = 0.15 \ \mathrm{fm^{-3}}$, binding energy per nucleon of $E_0/B = -15.2$ MeV, symmetry energy of $S_0 = 31.9$ MeV, symmetry energy slope of $L = 57$ MeV, and incompressibility of $K_0 = 267$ MeV \cite[e.g.,][]{PhysRevC.101.034904,Most:2022wgo}. The model can reproduce neutron stars heavier than $2 M_\odot$ and stars with sizes within the limits established by LIGO-Virgo \cite{PhysRevX.9.011001} and NICER  \cite{2019ApJ...887L..24M}.

\subsection{Neutron Star equation of state}

\medskip\noindent

The neutron star equation of state (EOS) we use is based on the Chiral Mean-Field (CMF) model presented in \cite{PhysRevC.101.034904}. A crust is simulated using \cite{1971ApJ...170..299B} by finding the intersection of the EOSs and merging them to determine the surface. This model has been chosen to describe our baryonic matter primarily because it is effective in producing neutron stars with masses and radii close to those observed by the LIGO/VIRGO collaborations, such as \cite{PhysRevX.9.011001}, and by NICER, \cite{2019ApJ...887L..24M}. Before the introduction of dark matter, the maximum neutron star mass with this model is $2.15$M$_\odot$ with a radius of $12.0$km. 
\\This, however, is not the only choice of EOS that could have been made, as countless viable models exist in the literature, for example the induced surface tension (IST) model,  \cite{2019ApJ...871..157S}, or indeed other relativistic mean field approaches, e.g. \cite{1997IJMPE...6..515S}.

\medskip\noindent

\subsection{Dark Matter  equation of state}

\medskip\noindent

Here, we introduce a relatively simple bosonic dark matter model, with an effective Lagrangian of the form
\begin{equation} \label{eq: LDM}
    \mathcal{L} =-\frac{1}{2} \partial_\mu\phi\partial^\mu\phi - \frac{1}{2} m_\chi^2 \phi^2 - \lambda \phi^4.
\end{equation}
where $\lambda$ relates to the scattering length, $l_\chi$, by $\lambda = \frac{4 \pi l_\chi}{m_\chi}$
\\A coherent wavefunction for the condensate can then be produced by considering the following ansatz,
\begin{equation}
\psi = \phi e^{im_\chi t},   
\end{equation}
from which the Lagrangian becomes
\begin{equation} \label{eq: Lphi}
    \mathcal{L} =-\frac{1}{2} g_{\mu\nu} \partial^\mu\psi^* \partial^\nu\psi - \frac{m_\chi ^2}{2}|\psi|^2 - \lambda|\psi|^4.
\end{equation}
By considering that the object rotates sufficiently slowly that the rotational motion of the condensate may be neglected, then we receive the following equations of motion, 
\begin{equation}\label{eq: Lpsi}
    i \frac{\partial \psi}{\partial t } = -\frac{1}{2 m_\chi} \nabla^2 \psi + m_\chi V \psi + \frac{4  \pi l_\chi}{m_\chi ^2} |\psi|^2 \psi
\end{equation}
\begin{equation}
    \nabla^2 V= 4 \pi G |\psi|^2
\end{equation}
known generally as the Gross-Pitaevskii-Poisson Equations,
where $V$ is the gravitational potential.
Following the methodology in \citet{2007JCAP...06..025B}, one sees that for an interaction term of the form found in equation \eqref{eq: Lphi}, and by identifying that $|\psi|^2\sim\rho$ ,the dynamics of the condensate reduce to that of a poltrope with the following equation of state:
\begin{equation} \label{eq:DMEOS}
	P_{r\chi} = A_\chi\; \rho_{\chi}^2,
\end{equation}
where
\begin{equation} \label{eq: A}
A_\chi =\frac{\lambda}{2 m_\chi^2} = \frac{2 \pi l_\chi}{m_\chi^3}=1.25\;10^5
\left(\frac{l_\chi}{1\;{\rm fm}}\right)
\left(\frac{1\; {\rm GeV}}{m_\chi}\right)^3,
\end{equation}
for which $m_\chi$ represents the mass of a dark matter particle, and $l_\chi$ denotes its scattering length. Throughout this work, including all tables, figures, and the main text, the units for $ A_\chi$ are presented in the cgs system unless specified otherwise: 	
${\rm g^{-1} \; cm^5 \; s^{-2}} $.
In this case, the dark matter condensate's equation of state is given by $P (\rho) $, as provided by equation \eqref{eq:DMEOS}. From this equation of state, it follows that $P_{r\chi} \propto \rho_{\chi}^2$. A general polytropic equation of state can be written as $P \propto \rho_{\chi}^{1 + 1/n}$, where $n=1$ is the polytropic index. 
 For the time being we only concern ourselves with the particular values of $A_\chi$ and the impact that this will have on neutron star structure.
\\Later we will consider the implications of this on the possible range of masses that dark matter may take on these regimes.
More detailed information about this equation of state in the context of dark matter may be found in \citet{2012JCAP...06..001L}, \citet{2015PhRvD..92d3011H}, \citet{2017PhRvD..96b3002P}, and \citet{2020EPJC...80..735C}.

\subsection{Anisotropic Dark Matter}
\begin{figure}[!t] 
\centering 
\includegraphics[width=1.1\columnwidth]{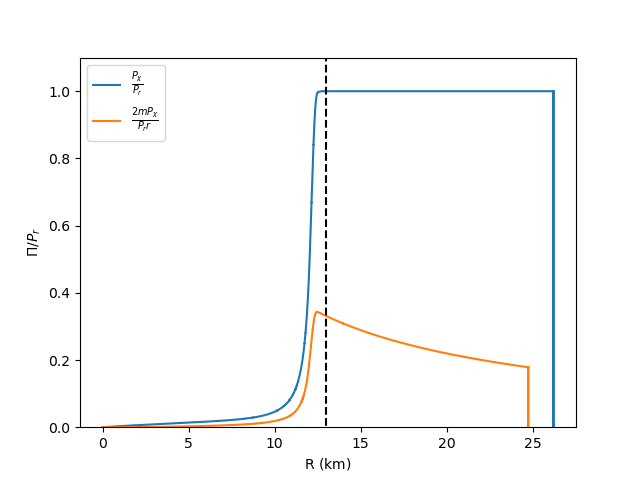}%
	\label{plot:sval}
\caption{The relative anisotropy of the system $\frac{2\Pi_\chi}{P_r} (r)$ (where $\Pi_\chi$ is given by equation \eqref{eq:Pi}) as a function of radius for $\alpha_\chi = 1$, $A_\chi = 10^7$, and for an object of total mass $M \sim 1.5M_\odot$. $\Pi_\chi (r)$ is computed from the center of the neutron star to the total radius $R_\chi$ of the halo for the two forms of anistotropy considered in equations \eqref{eq:mu1} and \eqref{eq:mu2}. The dashed line indicates the very edge of the neutron star, located at approximately $13.0$ km, with $\frac{2\Pi_\chi}{P_r} (r)$ peaking for both models at around $12.6$ km.
}
\label{fig:Pichi}
\end{figure} 

\medskip\noindent
Anisotropy in stars can originate from various sources. For instance, boson stars are inherently anisotropic astronomical objects \citep{2003CQGra..20R.301S}, and the energy-momentum tensors of both electromagnetic and fermionic fields are intrinsically anisotropic \citep{2004IJMPD..13.1389G}. Local anisotropy could arise from the presence of viscosity \citep[e.g.,][]{1993Ap&SS.201..191B} or from a two-fluid mixture \citep[e.g.,][]{1980PhRvD..22..807L,1982PhRvD..26.1262B,2022PhRvD.106l4023N}. In this context, several authors have recently developed stellar models incorporating anisotropic matter \citep{1975A&A....38...51H,2023PhRvD.107f4010L,2023Ap&SS.368...21R}.

\medskip\noindent
This study aims to investigate the consequences of introducing anisotropy into a dark matter model. As mentioned earlier, the motivation for including such an effect can be attributed to the bosonic nature of dark matter particles, which inherently exhibit anisotropy.

\medskip\noindent
 A viable approach to incorporating anisotropy in the dark matter model has been explored by \citet{2011CQGra..28b5009H} and others \citep{1974ApJ...188..657B,2015PhRvD..91d4040F}. In this approach, the anisotropy function $\Pi_\chi(r)$ is defined as:
\begin{equation}
	\label{eq:Pi}
	\Pi_\chi = P_{\perp\chi} - P_{r\chi} = \frac{\alpha_\chi \mu_\chi}{2} 
\end{equation}
Here, $\alpha_\chi$ denotes a dimensionless constant characterizing the strength of the anisotropy, while $\mu_\chi$ defines the relationship between system anisotropies and their local properties. We explore two different models for $\mu_\chi(r)$:
\begin{enumerate}	
\item
In the first model, commonly adopted in many articles, we assume that $\Pi_\chi$ is directly proportional to $P_{\chi r}$, resulting in the equation:
\begin{equation}\label{eq:mu1}
	\mu_\chi(r) = P_{\chi r}
\end{equation}
\item
In a more recent description found in the literature, the anisotropy is also proportional to the total enclosed mass $m(r)$ inside a sphere of radius $r$. Thus, we have:
\begin{equation}\label{eq:mu2}
	\mu_\chi(r) = \frac{2m(r)}{r} P_{\chi r}.
\end{equation}
Here, $m(r)$ represents the combined mass, including both dark matter and baryonic matter, within a sphere of radius $r$.
The choice of equation \eqref{eq:mu2} for the anisotropy factor has several appealing features.
First, the compactness $ \mu_\chi $ scales approximately as $ r^2$, while $m(r)$ scales roughly as $ r^3$ as $ r $ approaches zero. This leads to a vanishing anisotropy at the centre, guaranteeing the regularity of the right-hand side of equation \eqref{eq:TOVi} when $ i=\chi$.
Second, the anisotropy factor becomes significant only for highly relativistic configurations where $\mu_\chi \sim \mathcal{O}(1)$. This is consistent with the consensus that fluid anisotropy increases at higher matter densities.
\end{enumerate}	
The realisation of both models may be seen in Fig. \ref{fig:Pichi}, where we consider the size of the anisotropy relative to the total radial pressure $P_r$. For equation \eqref{eq:mu1} we see that the relative anisotropy, $\frac{2\Pi_\chi}{P_r}$, increases smoothly until it reaches the edge of the neutron star, after which the anisotropy is equivalent to the total radial pressure, as there is no baryonic pressure outside of the neutron star.
\\In the regime governed by equation \eqref{eq:mu2}, the relative anisotropy increases at a lower rate, and once outside of the neutron star it instead now slowly declines until it reaches the edge of the dark matter halo.
\\One can see that despite both objects having identical initial conditions the latter model produces halos of smaller extent.



\section{Modelling Dark Matter Admixed Neutron Stars}
\label{sec:DNR}
\begin{figure}[!h]
	\centering
	\includegraphics[width=1.1\columnwidth]{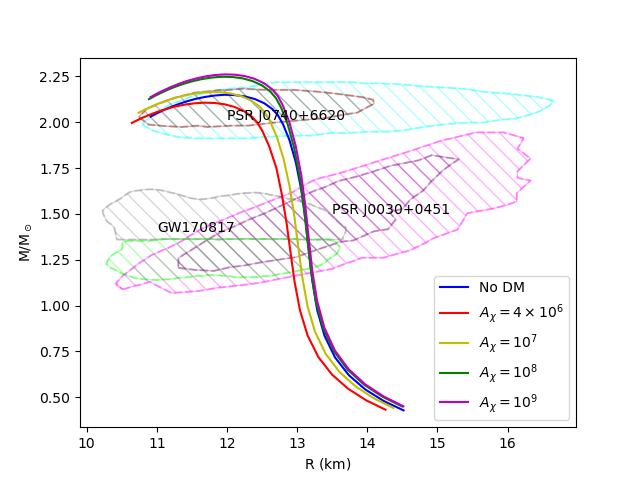}
	\caption{Results of hybrid neutron star simulations for all choices of $A_\chi$ before the inclusion of anisotropy, $\alpha_\chi= 0$. Included also is the contour data pertaining to neutron star mass and radii as measured by NICER and LIGO. For PSR J0740+6620, a 68\% credibility region is highlighted in maroon, and 95\% credibility region in turquoise,\cite{2021ApJ...918L..27R} and \cite{2021ApJ...918L..28M}.For PSR J0030+0451, the purple region indicates 68\% credibility and the fuchsia region 95\%, \cite{2019ApJ...887L..24M}. For GW170817, the upper grey region and lower green region indicate the major and minor component respectively of the event, \cite{2018PhRvL.121p1101A}.}
	\label{fig:allAalph0}
\end{figure}

\subsection{Hybrid Neutron Star Simulations}

\medskip\noindent
In the computation of these hybrid stars, which consist of a neutron star and a dark matter halo, we have made several numerical adjustments to ensure the accuracy and reliability of our results. Specifically, we highlight the following considerations:
\begin{enumerate}
\item	
The equation of state for a neutron star exhibits a hard edge, meaning that numerical tables do not smoothly approach zero pressure or density. Therefore, determining the edge of the object accurately from a numerical perspective is crucial, as selecting a region where pressure or density is zero lies outside the scope of an equation of state interpolated from pressure-density data.
\\As such, the surface of the neutron star is determined as the radius at which the increase in mass due to baryonic components produced by increasing the radius by one metre is less than a certain threshold. 

\item
We must similarly be careful in considering the properties and size of the dark matter halo. The equation of state is smooth all the way to zero pressure and density, and as such we consider the dark matter halo ending at a radius $R_\chi$ where $P_{r\chi}(R_\chi) = 0$.
\\Depending on the value of $A_\chi$, the dark matter halo has the capacity to become extremely large, meaning that we may be concerned with its interaction with other astrophysical objects. The maximum extent of a Bose-Einstein Condensate (BEC) halo of this kind will essentially be its Lane-Emden radius, as governed by the Lane-Emden equations for a polytropic fluid which may be seen in \cite{2020EPJC...80..735C} in a similar context.
\\From this we see that the Lane-Emden radius is described by 
\begin{equation}
	R_{LE} = \sqrt{\frac{\pi A_\chi}{2}},
\end{equation}
which for $A_\chi = 10^{10}$ would produce a radius of around $4800$km, and therefore the domains probed in this article should be free from any of the aforementioned concerns.

\end{enumerate}
\medskip\noindent
For each parameter used, we simulate a set of 30 hybrid neutron stars with masses ranging from 0.4$M_\odot$ to 2.4$M_\odot$. These simulations are specified by defining separate central pressures and densities for both the dark and baryonic matter. Subsequently, numerical integration is performed outward to a specified radius, chosen to ensure that the full extent of the dark matter halo is completely encompassed.
\\All models produced have a dark matter fraction of $5\%$ of the neutron star mass, and this parameter will be kept fixed across all simulations so that relevant differences produced by other properties may be effectively identified. We choose this fiducial value as it is in line with quantities seen in the wider literature, e.g. \cite{2020PhRvD.102f3028I}, as well as that the effects and differences between models are felt to be sufficiently visible for this choice.

\subsection{Properties of the Dark Matter Halo}
\label{sec:DMH}  

\medskip\noindent
In this study, we explore the behaviour and properties of dark matter halos described by a variety of choices of $A_\chi$, and effected by several levels of anisotropy as dictated by $\alpha_\chi$. We will therefore be able to see the effects of different choices of dark matter equation of state properties, the effects of introducing anisotropy, as well as how significant the manifestation of an anisotropy is in each regime.

\subsubsection{The Equation of State Parameter $A_\chi$}

\begin{table}[]
	\centering
	\begin{tabular}{lllllll}
		\hline
		$A_\chi$& $M_{max}$ & $R_B(M_{max})$ & $R^{min}_{B}$&$R^{max}_{B}$& $R^{min}_{\chi}$& $R^{max}_{\chi}$\\ 
  		${\frac{{\rm cm^5}}{\rm gs^2}}$& M$_\odot$ & km &km & km & km &  km\\ 
  		\hline\hline
		$0$& $2.15$& $12.0$ & $10.9$&$14.5$   &  & \\ 
		$4\;10^6$& $2.11$ & $11.7$ &$10.6$&$14.3$  & $12.5$&$17.8$ \\ 
		$10^7$& $2.17$ & $11.8$ &$10.7$&$14.4$  & $20.2$&$26.9$ \\ 
		$10^8$& $2.25$ & $11.9$ &$10.9$&$14.5$  & $77.4$&$84.2$ \\ 
		$10^9$& $2.26$ & $12.0$ &$10.9$&$14.5$  & $262.1$&$268.2$ \\ 
	\end{tabular}
\caption{Mass-Radius data for hybrid compact objects containing isotropic BEC dark matter for various values of $A_\chi$, for which $\alpha_\chi = 0$. Within we observe the maximum mass seen in figure \ref{fig:allAalph0}, the associated baryonic radius, as well as the range of baryonic and dark radii observed'}
\label{tab:my-table}
\end{table}	

\medskip\noindent
We now begin by discussing the effect of varying $A_\chi$, the constant controlling our polytropic equation of state, equation \eqref{eq:DMEOS}. As one may see in equation \eqref{eq: A}, these choices will also be indicative of the range of masses that these models will represent, depending on the choice of scattering length that is made.

\medskip\noindent
The nuances of these $A_\chi$ values, before factoring in anisotropy, are illustrated in Figure \ref{fig:allAalph0} and enumerated in Table \ref{tab:my-table}. The table delineates mass-radius data for hybrid compact objects embedded with isotropic Bose-Einstein Condensate (BEC) dark matter. The data span a range of $ A_\chi $ from $ 4 \; 10^6 $ to $ 10^9$, with all observations adhering to the $ \alpha_\chi = 0 $ stipulation. For completeness we include the smallest and largest radii simulated, denoted $R^{min}$  and $R^{max}$ respectively. It is worth noting the smallest dark radius belongs to the smallest baryonic radius neutron star, and the largest dark radius to the largest baryonic radius object.

\medskip\noindent
When juxtaposing the data from Table \ref{tab:my-table} with Figure \ref{fig:allAalph0}, two distinct behaviours in the mass-radius relations of these hybrid objects emerge based on the type of dark matter: (i) objects with denser dark matter (indicated by smaller $A_\chi$) appear more compact at lower masses and resemble characteristics of non-dark matter objects at higher masses; (ii) on the other hand, with more diffuse dark matter (larger $A_\chi$), neutron stars at lower masses are akin to those without dark matter, but display notably larger masses at reduced radii as their masses increase.

\medskip\noindent
Delving deeper into Table \ref{tab:my-table}, we note that when $A_\chi = 4\;10^6$, it records the smallest radius across all parameters and also the least maximum mass. As the value of $A_\chi$ escalates, there is a noticeable surge in the dark matter halo radius, with the most larger dark radii observed at $ A_\chi= 10^9$. Meanwhile, the baryonic radius seems to stabilize for values of $ A_\chi > 10^8 $, showcasing a range of radii akin to a conventional neutron star. The masses of all the examined neutron stars peak between $11.5$ km and $12$ km. Both $A_\chi = 10^8$ and $10^9$ possess strikingly similar maximum masses, notably around 5\% more than $ A_\chi = 0 $. 

\begin{figure}[!h] 
	\centering 
	\subfloat{%
		\includegraphics[width=1.1\columnwidth]{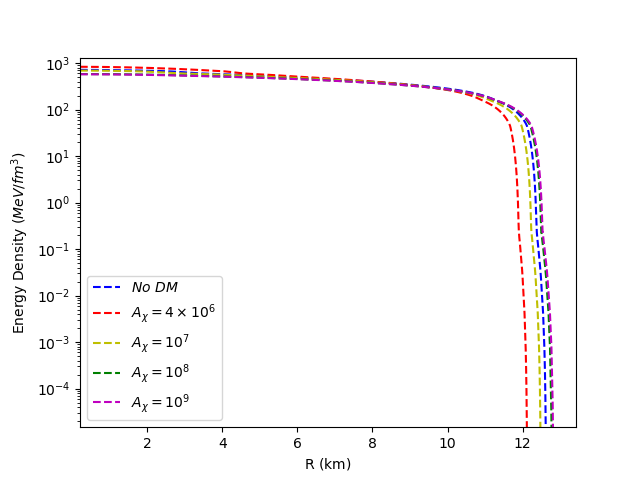}%
		\label{plot:BaryR}%
	}\qquad
	\\
	\subfloat{%
		\includegraphics[width=1.1\columnwidth]{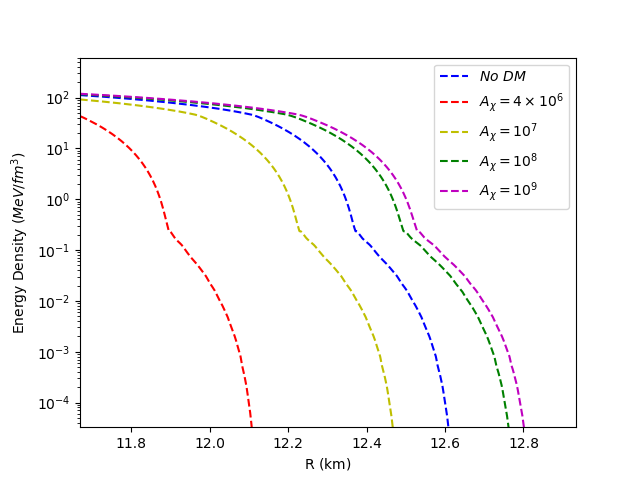}%
		\label{plot: BaryRz}%
	}
\caption{
The figure shows the Energy Density-Radius curves for the baryonic matter present in neutron stars with a mass of $2.083$M$_\odot$ for different values of $A_\chi$.
For diminished $A_\chi$ values, the baryonic radius is notably smaller than in cases devoid of dark matter, revealing pronounced differences in radii. In contrast, for augmented $A_\chi$ values, the baryonic radii increase and converge. The lower panel provides a detailed view of the Density-Radius profiles near the neutron star's radius.}
	\label{fig:BaryR}
\end{figure} 	
\medskip\noindent
In Figure \ref{fig:BaryR}, we further explore this analysis, where we showcase the density-baryonic radius relationship of five compact objects of similar mass but distinct compositions. For $ A_\chi =  4\;10^6 $ and $ 10^7 $, the object is more compact than a standard neutron star, evidenced by a denser inner core and reduced radius. Objects with $A_\chi = 10^8 $ and $ 10^9 $ have radii that align more with conventional neutron stars, leading to the speculation that heavier dark matter might cause the baryonic matter to be more compact compared to a neutron star. Meanwhile, larger-radius, lighter dark matter seems to expand the baryonic radius beyond its typical dimensions.

\medskip\noindent
However Figure \ref{fig:BaryR} does not tell the full story.
Each hybrid object has a baryonic mass approximately equal to $1.98 $ M$_\odot$, as opposed to its total combined mass of $2.083 $ M$_\odot$, a value that for a standard neutron star generally equates to a radius of 
$ 12.81$ km. This suggests that in all objects mixed with dark matter, the baryonic matter is more densely packed. For greater values of $ A_\chi $, the baryonic radius is in close proximity to a non-hybrid object of comparable baryonic mass. It suggests that lighter dark matter has a more limited impact on the baryonic structure of the neutron star and mainly contributes to the overall dark mass, and in essence a $1.98 $ M$_\odot$ neutron star appears to have a dark matter cloud superimposed on top of it.

\begin{figure}[!h] 
	\centering 
	\subfloat{%
		\includegraphics[width=1.1\columnwidth]{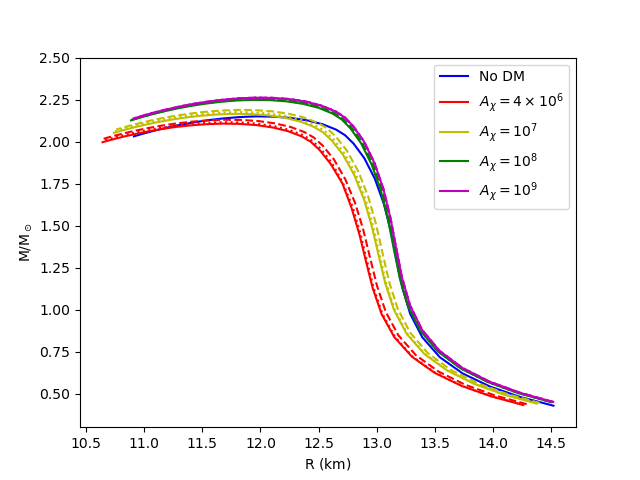}%
		\label{plot:anisoalls}%
	}\qquad
	\\
	\subfloat{%
		\includegraphics[width=1.1\columnwidth]{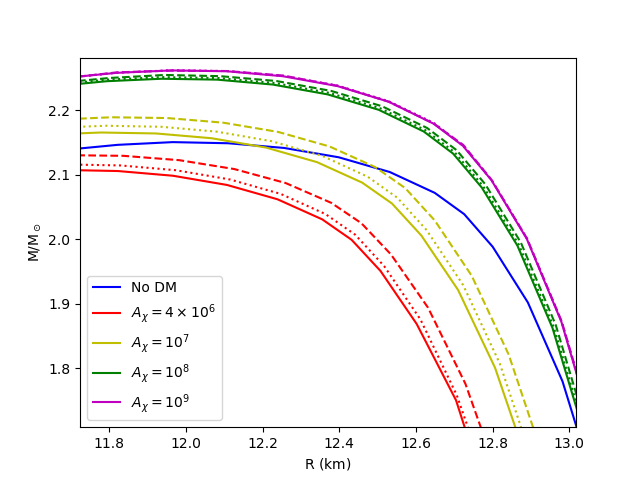}%
		\label{plot:anisoallszoom}%
	}
	\caption{Plots containing mass-radius relations for all choices of $A_\chi$ and $\mu_\chi$, where we also have $\alpha_\chi = 1$. Solid lines indicate curves without anisotropy, dotted lines indicate the presence of anisotropy governed by Eq. \ref{eq:mu1}, and dashed lines an anisotropy governed by Eq. \ref{eq:mu2}.  }
	\label{fig:MRanisos}
\end{figure} 

\subsubsection{Variations in Anisotropy Models}

\medskip\noindent	
We now proceed to examine the influence of introducing anisotropy into our dark matter. Furthermore, we contrast the disparities between the anisotropy models described by equations \eqref{eq:mu1} and \eqref{eq:mu2} and observe their behaviours for each dark matter variant under consideration. In Figure \ref{fig:MRanisos}, the mass-radius relations corresponding to all $A_\chi$ values are presented: both in the absence of anisotropy and when anisotropies are governed by equations \eqref{eq:mu1} and \eqref{eq:mu2}, with $\alpha_\chi = 1$.
It is evident that the mass-radius curves generated by equation \eqref{eq:mu2} (represented by dashed lines) possess higher masses for a specific radius than those from equation \eqref{eq:mu1}. This distinction becomes most pronounced when considering lower values of $A_\chi$.

\medskip\noindent
The contrast between models with and without anisotropy diminishes with increasing $A_\chi$ values, to the extent that all three curves become virtually indistinguishable at $A_\chi = 10^9$. Recalling prior discussions, this trend can likely be attributed to the need for a decreased central pressure to form a dark matter halo constituting $5\%$ of the aggregate mass for elevated $A_\chi$ values. Consequently, the radial contribution to pressure within the neutron star from dark matter is substantially overshadowed by that of the baryonic matter. This renders the TOV equations largely unresponsive to the anisotropies inherent in the dark matter. For $A_\chi = 10^9$, the curve closely parallels the one devoid of dark matter, with each mass incrementally increased by roughly $5\%$. This implies that, in this scenario, the dark matter primarily serves to augment the object's mass without significantly influencing the baryonic radius. As a result, the anisotropic nature of the matter exerts negligible influence the mass-radius relation.

\subsection{Impact of Anisotropy on Halo Structure}

\medskip\noindent
We delve into the influence of anisotropies on the structural characteristics of the dark matter halo and investigate the resultant shifts in mass-radius curves as $\alpha_\chi$ transitions from 0 to 1.
\begin{figure}[!h] 
	\centering 
	\subfloat{%
		\includegraphics[width=1.1\columnwidth]{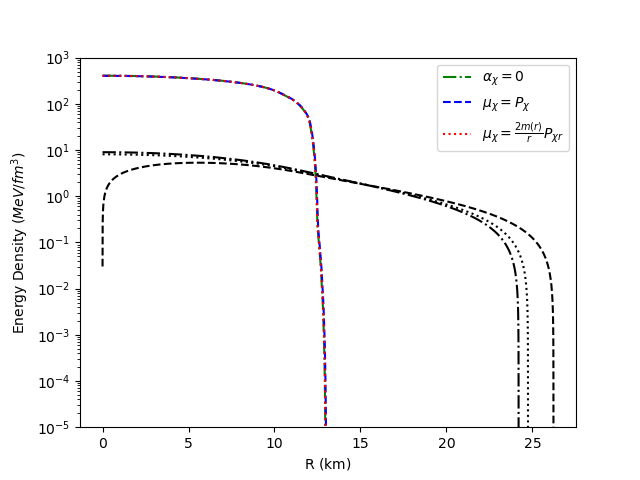}%
		\label{plot:anisoalls}%
	}\qquad
	\\
	\subfloat{%
		\includegraphics[width=1.1\columnwidth]{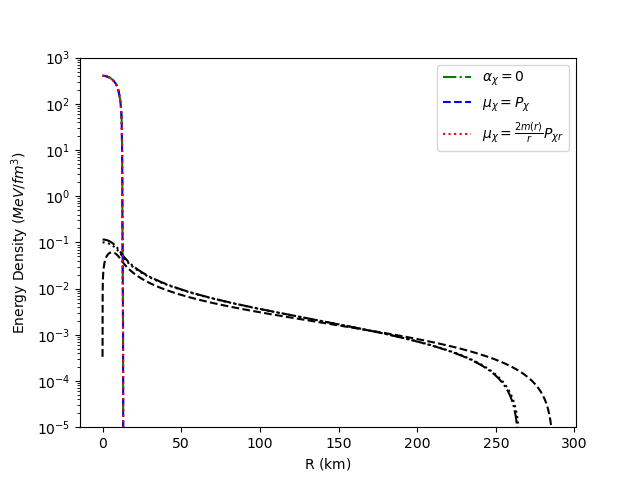}%
		\label{plot:anisoallszoom}%
	}
	\caption{The Energy Density-Radius curves for dark matter admixed neutron stars with and without anisotropy. For $A_\chi = 10^7$ (upper) we consider $1.50$M$_\odot$ objects, and for $A_\chi = 10^9$ (lower) $1.54$M$_\odot$ objects. The anisotropies are governed by equations \eqref{eq:mu1} (dashed) and \eqref{eq:mu2} (dotted) where $\alpha_\chi = 1$, as well as including $\alpha_\chi = 0$ (dot-dash) as a reference. Baryonic matter is indicated by a coloured line, and the black line of the respective line style indicates the associated dark matter curve.}
	\label{fig:DRcurves}
\end{figure} 
\medskip\noindent
Figure \ref{fig:DRcurves} illustrates the repercussions of each anisotropy model on the halo's structure for $A_\chi=10^7$ and $A_\chi=10^9$. The impact of the former can be discerned in the upper panel of figure \ref{fig:DRcurves}. The halo with the smallest size arises in the absence of anisotropy, while the dynamics governed by the equation \eqref{eq:mu1} mould the largest halo.
Notably, the behaviour steered by the anisotropies of equation \eqref{eq:mu1} deviates notably from the patterns seen in equation \eqref{eq:mu2} and the non-anisotropic case near $r=0$. Even though anisotropies are often modelled in this manner (as found in the literature), the resultant attributes at the core of the star might be deemed unrealistic, differentiating it from the alternative model (equation \ref{eq:mu2}) presented in this analysis.

\medskip\noindent
All dark matter halos have uniform masses; therefore, halos governed by anisotropies directly proportional to pressure inevitably exhibit larger sizes to offset their less dense cores. For anisotropies governed by equation  \eqref{eq:mu2}, the changes are more subtle. Within the neutron star, the dark matter distribution exhibits a slightly reduced density compared to its isotropic counterpart but spans an additional few hundred meters in radius. This pattern persists in the lower panel of Figure 
 \ref{fig:DRcurves}, with the disparities between the two models amplifying. Remarkably, while equation \eqref{eq:mu2} engenders a halo only marginally larger than the isotropic case, equation \eqref{eq:mu1} continues to feature its hallmark low-density core. Moreover, the halo influenced by equation \eqref{eq:mu1} now stretches several tens of kilometers further than the other two scenarios.

\section{Constraints on the Properties of Dark Matter Particles}
\label{sec:PDMP}  

\medskip\noindent

In light of our investigation into these bosonic dark matter models, we will now discuss potential constraints on the dark matter properties in the context of astrophysical observations. While this study primarily focuses on a hybrid dark matter-neutron star model considering only gravitational interactions, we also explore dark matter models wherein the self-interaction of dark matter and its interaction with baryons are so negligible that they can be approximated as non-interactive at a first approximation.
\\As our dark matter EOS is fully specified by a choice of $A_\chi$, we have been free to simulate hybrid neutron stars without the need to concern ourselves with more specific properties of the dark matter particles. With a known value of $A_\chi$, equation \ref{eq: A} can be used to fully specify the particle mass if its scattering length is known, and vice versa. As such we will now consider possible constraints that may be set on our system based on both astrophysical data, and terrestrial experiments on BEC gases.
\vspace{1cm}
\subsection{Galactic Cluster Constraints on Sub-GeV Bosonic Dark Matter Particle Masses}

\begin{table}[h]
	\centering
	\begin{tabular}{llll}
		\hline
		$A_\chi$&  $l_\chi$& $m_\chi$\\
		${\frac{{\rm cm^5}}{\rm gs^2}}$& fm &  GeV\\ 
		\hline \hline
		$4\;10^6$& $2.79$ & $0.442$   \\ 
		$10^7$& $2.33$ & $0.306$ \\ 
		$10^8$& $1.47$ & $0.122$ \\ 
		$10^9$& $0.93$ & $0.0485$\\ 
	\end{tabular}
	\caption{Properties of dark matter corresponding to $\sigma^{bc}_m = 1.25$ cm$^2/$g, tabulated against various values of the polytropic constant $A_\chi$.}
	\label{tab:my-table2}
\end{table}

\noindent
In this study, we refine our bosonic dark matter model using contemporary astronomical observations. Notably, collision data from galaxy clusters, such as the Bullet Cluster \citep[1E 0657-56,][]{2004ApJ...604..596C} and the Baby Bullet \citep[MACSJ0025-12,][]{2008ApJ...687..959B}, provide essential insights into the properties of dark matter.
Further observations of galaxy cluster collisions, notably MACS J0025.4-1222 \citep{2008ApJ...687..959B}, reveal mismatches between the centers of visible matter and gravitational mass. \citet{2006ApJ...648L.109C} determined that the observed absence of dark matter deceleration offers a constraint on its self-interaction strength. Specifically, this can be expressed in terms of the self-interaction cross section for long-range forces \citep[e.g.,][]{2017FrPhy..12l1201Y}. This observational data provides constraints on boson dark matter particles, represented by the scattering cross section $\sigma_\chi$. This cross section  is directly related to the scattering length according to $\sigma_\chi=4\pi l_\chi^2$.  

\medskip\noindent
Under the $\Lambda$CDM cosmological model, based on galaxy merger observations, $\sigma_m$ should not exceed a critical value $\sigma^{bc}_m$: $\sigma_m\le \sigma^{bc}_m$. Studies, such as \citet{2020EPJC...80..735C} and \citet{2015MNRAS.452.3030T}, cite the Bullet Cluster and Baby Bullet as evidence for this upper limit on the ratio $\sigma_m$.
\citet{2008ApJ...679.1173R} found that $\sigma^{bc}_m = 1.25$ cm$^2/$g denoting the maximal value consistent with these findings, we can deduce constraints on the dark matter particle mass and scattering length for each of our dark matter proposed model, ensuring their coherence with this astronomical observation.

\medskip\noindent
We initiate our analysis with the ratio:
\begin{equation}
	\sigma_m = \frac{\sigma_\chi}{m_\chi}=\frac{4 \pi l_\chi^2}{m_\chi}.
\end{equation}
We can derive constraints on $m_\chi$ and $l_\chi$. Utilizing $\sigma^{bc}_m = 1.25$ cm$^2/$g:
\begin{equation}
	\left(\frac{m_\chi}{\rm 1 \; GeV} \right) =0.0565
	\left(\frac{l_\chi}{\rm 1\; fm}\right)^2.
\label{eq:mlchi}	
\end{equation}

\medskip\noindent
Here, we aim to find parameters to our  the validity of our model. Considering that observational constraints predominantly hinge on the scattering properties of galactic dark matter, and in alignment with this study's objectives, we rewrite equation \eqref{eq: A} in terms of the mass of the dark matter particle:
\begin{equation}
A_\chi = \frac{\sqrt{\pi\sigma_c^{bc}}}{m_\chi^{5/2}}.
\label{eq:Amchi}	
\end{equation}
By setting this and considering $\sigma_m=\sigma^{bc}_m$, we can deduce the parameters associated with the dark matter's mass and scattering length. Table \ref{tab:my-table2} presents the properties of dark matter, with each row detailing the associated values of the polytropic constant $ A_\chi$, scattering length $ l_\chi $, and mass $ m_\chi $ for a given dark matter configuration corresponding to $ \sigma^{bc}_m = 1.25 $ cm$^2$/g. 
Using the algorithm based on  $A_\chi $, we determine 
$m_\chi$ and  $l_\chi$ with equations \eqref{eq:Amchi} and \eqref{eq:mlchi}.
The results indicate that, within the scope of $A_\chi$ addressed in this study, $m_\chi$ ranges from $0.05$ GeV to $0.5$ GeV, while $l_\chi$ lies between $0.9$ fm and $3$ fm.
These constraints are particularly intriguing because this mass range of bosonic dark matter has not been extensively discussed in the literature. In fact, much of the literature \citep[e.g.,][]{2020EPJC...80..735C} primarily emphasizes lightweight BEC dark matter species with significantly smaller masses, associated with dark matter halos spanning kilo-parsec radii.

\subsection{Experimental Constraints on Sub-GeV Bosonic Dark Matter Particle Masses}

\begin{table}[h]
	\centering
	\begin{tabular}{ll}
		\hline
		$A_\chi$& $m_\chi$ \\ 
		${\frac{{\rm cm^5}}{\rm gs^2}}$& GeV  \\
		\hline \hline
		$4\;10^6$& $31.5$   \\  
		$10^7$& $23.2$  \\ 
		$10^8$& $10.8$ \\ 
		$10^9$& $5.0$  \\ 	
	\end{tabular}
	\caption{Properties of dark matter corresponding to $l^{ex}_\chi = 10^6\;{\rm fm}$, tabulated against various values of the polytropic constant $A_\chi$.}
	\label{tab:my-table3}
\end{table}

\medskip\noindent
A Bose-Einstein Condensate (BEC) typically forms when the temperature of a system drops beneath the critical temperature, represented as:
\begin{equation}
	T_c = \frac{2\pi}{m_\chi}
	\left(\frac{n_\chi}{\zeta(3/2)}\right)^{2/3},
\end{equation}
where $ n_\chi = \frac{\rho_\chi}{m_\chi}$ and $\zeta(3/2)$ is the Riemann zeta function. This critical temperature, $T_c $, is characterized by the thermal de Broglie wavelength equating to the mean interparticle distance, facilitating the synchronization and overlap of individual particle wave functions \citep[e.g.,][]{2014EPJD...68..341G}. A coherent state emerges when particle density is sufficiently high or the temperature is adequately low
\citep[e.g.,][]{1999RvMP...71..463D}.
Given the properties of BEC dark matter, its interactions with the observable universe are predominantly gravitational. Hence, we hypothesized that the temperature of the condensate will experience negligible alterations upon accretion onto a neutron star. Consequently, we assume that the dark matter temperature stays below the critical temperature  $T_c$.

\medskip\noindent
Bose-Einstein Condensation (BEC) is a well-established phenomenon in terrestrial experiments \citep[e.g.,][]{1999RvMP...71..463D}. Therefore, the idea that similar condensation could occur elsewhere in the universe is not implausible \citep[e.g.,][]{2012PhRvD..86f4011C,2014EPJD...68..341G,2020EPJC...80..735C}. Specifically, if the temperature of a boson gas falls below its critical temperature, BEC can manifest at some point in the universe's cosmic history. Indeed, terrestrial experiments with cold gases have already verified the presence of this unique state of matter for bosonic particles in ultra-cold, low-density conditions \citep{1995Sci...269..198A,PhysRevLett.75.3969}.

\medskip\noindent
Now, we turn our focus to the scattering lengths observed in terrestrial Bose-Einstein Condensate (BEC) gas experiments, as documented in \cite{PhysRevLett.75.3969} and \cite{PhysRevLett.74.5202}.
For instance, experimentally measured scattering lengths are typically 2.75 nm for $^{23}Na$
 \citep{141041} and 5.77 nm for 
 $^{87}Rb$ \citep{1997PhRvA..55..636B}. The use of constraints set by considering ordinary matter provides a counterpoint to those set by astronomical observations. As the precise nature of dark matter is not yet understood, it is valuable for us to consider the implications of bosonic dark matter with properties more similar to condensates encountered terrestrially.
 \\If there exists a zoo of dark matter particles, much like with the standard model, then it can be justifiable to explore the possibility that the accumulated dark matter is distinct from that make up galactic halos.
 \\For our calculations, we adopt a benchmark scattering length, $l^{ex}_\chi =10^6$ fm, which aligns well with contemporary laboratory findings.   Aligned with the objectives of this study, we recast equation \eqref{eq: A} to be articulated in terms of the dark matter particle's mass:
 \begin{equation}
 	A_\chi = \frac{2\pi l^{ex}_\chi}{m_\chi^3}
 	= 1.25\;10^{11}\; \left(\frac{1\;{\rm GeV}}{m_\chi}\right)^{3},
 \end{equation}
 taking into consideration our predefined value for $l^{ex}_\chi$. Using this formula, we can determine the dark matter's mass parameters. Table \ref{tab:my-table3} lists these values, showing the corresponding polytropic constant  $A_\chi$
 and mass  $m_\chi$.

 
 \section{Summary and Conclusion} 
\label{sec-Con}
 
\medskip\noindent
 In this study, we examined the intricacies surrounding the properties of dark matter particles using experimental data and astrophysical observations, with a special emphasis on the hybrid dark matter-neutron star model.  
 
 \medskip\noindent
 We initiated our analysis by modelling a hybrid neutron star embedded within an anisotropic bosonic dark matter halo without pre-defined constraints on the dark matter particles' mass range or scattering length. Varying the equation of state parameter $A_\chi$, we derived distinct mass-radius correlations for these hybrid constructs. Notably, for our fiducial dark matter model, the parameter $A_\chi$ needs to lie between $4 \; 10^6$ and $10^9$ g$^{-1}$ cm$^5$ s$^{-2}$ to profoundly influence the empirically observed mass-radius relationship of the neutron star.
It became evident that the dark matter's nature, whether condensed or dispersed, has a significant bearing on the characteristics of neutron stars correlated with their masses. Introducing anisotropies further diversified the mass-radius trajectories. Among the models assessed, equation \eqref{eq:mu2} emerged as an authentic representation of anisotropy.  This model is distinct in its generation of halos with a unique low-density core profile, distinguishing it from alternative model forecasts. 
These anisotropic nuances influence the dark matter halo's intrinsic properties, especially in size, density gradient, and the neutron star's baryonic component.

\medskip\noindent
Galactic clusters, specifically the Bullet Cluster, have been instrumental in refining our bosonic dark matter model. From this analysis, we have inferred a relationship for bosonic dark matter particles, characterized by the scattering cross section $\sigma_\chi$, with the scattering length described as $\sigma_\chi=4\pi l_\chi^2$.This insight sharpens our understanding of the parameter $m_\chi$, which we have now restricted between 0.05 GeV to 0.5 GeV and $l_\chi$ in the range of 0.9 fm to 3 fm. Grounded in terrestrial Bose-Einstein Condensate (BEC) experiments and using a benchmark scattering length $l_\chi \approx 1$  nm, we delineated the constraints on dark matter properties. With a fiducial scattering length set at $l_\chi = 10^6 \; \text{fm}$, our model suggests dark matter particle masses to fall between 3 GeV and 30 GeV. 

\medskip\noindent
Notably,
the dark matter particle mass range identified in the preceding sub-sections, specifically between $10^2$ to $10^{-3}$ GeV, presents notable challenges for direct dark matter detection experiments \citep{2019ARNPS..69..137D}. Specifically, in the mass range from $10^{-3}$ to $0.5$ GeV, current detectors face technical limitations exacerbated by an anticipated substantial neutrino background
\citep{2014PhRvD..90h3510R}.

\medskip\noindent
In conclusion, our study illuminates the intricate relationship between dark and baryonic matter in hybrid neutron stars. The variability in outcomes, influenced by different equations of state and anisotropic models, underscores the imperative for more detailed investigations. Given the insights we have gleaned from galactic clusters and terrestrial BEC experiments, this work is a relevant reference point, guiding subsequent research into the enigmatic realm of dark matter and its astronomical manifestations.

\section*{Acknowledgments}
\medskip\noindent

I.L. and Z.B.S. gratefully acknowledge the financial support provided to the Center for Astrophysics and Gravitation (CENTRA/IST/ULisboa) through the Grant Project No. UIDB/00099/2020 and Grant No. PTDC/FIS-AST/28920/2017 by the Fundação para a Ciência e Tecnologia (FCT), Portugal.
\\We also gratefully thank the anonymous referee for their positive insights and suggestions, which have surely improved the quality of this work.

\newpage
\bibliography{artNSDM23}

\begin{thebibliography}{99}
\expandafter\ifx\csname natexlab\endcsname\relax\def\natexlab#1{#1}\fi
\expandafter\ifx\csname bibnamefont\endcsname\relax
  \def\bibnamefont#1{#1}\fi
\expandafter\ifx\csname bibfnamefont\endcsname\relax
  \def\bibfnamefont#1{#1}\fi
\expandafter\ifx\csname citenamefont\endcsname\relax
  \def\citenamefont#1{#1}\fi
\expandafter\ifx\csname url\endcsname\relax
  \def\url#1{\texttt{#1}}\fi
\expandafter\ifx\csname urlprefix\endcsname\relax\def\urlprefix{URL }\fi
\providecommand{\bibinfo}[2]{#2}
\providecommand{\eprint}[2][]{\url{#2}}

\bibitem[{\citenamefont{{Battaglieri} et~al.}(2017)\citenamefont{{Battaglieri}, {Belloni}, {Chou}, {Cushman}, {Echenard}, {Essig}, {Estrada}, {Feng}, {Flaugher}, {Fox} et~al.}}]{2017arXiv170704591B}
\bibinfo{author}{\bibfnamefont{M.}~\bibnamefont{{Battaglieri}}}, \bibinfo{author}{\bibfnamefont{A.}~\bibnamefont{{Belloni}}}, \bibinfo{author}{\bibfnamefont{A.}~\bibnamefont{{Chou}}}, \bibinfo{author}{\bibfnamefont{P.}~\bibnamefont{{Cushman}}}, \bibinfo{author}{\bibfnamefont{B.}~\bibnamefont{{Echenard}}}, \bibinfo{author}{\bibfnamefont{R.}~\bibnamefont{{Essig}}}, \bibinfo{author}{\bibfnamefont{J.}~\bibnamefont{{Estrada}}}, \bibinfo{author}{\bibfnamefont{J.~L.} \bibnamefont{{Feng}}}, \bibinfo{author}{\bibfnamefont{B.}~\bibnamefont{{Flaugher}}}, \bibinfo{author}{\bibfnamefont{P.~J.} \bibnamefont{{Fox}}}, \bibnamefont{et~al.}, \bibinfo{journal}{arXiv e-prints} \bibinfo{eid}{arXiv:1707.04591} (\bibinfo{year}{2017}), \eprint{1707.04591}.

\bibitem[{\citenamefont{{Billard} et~al.}(2022)\citenamefont{{Billard}, {Boulay}, {Cebri{\'a}n}, {Covi}, {Fiorillo}, {Green}, {Kopp}, {Majorovits}, {Palladino}, {Petricca} et~al.}}]{2022RPPh...85e6201B}
\bibinfo{author}{\bibfnamefont{J.}~\bibnamefont{{Billard}}}, \bibinfo{author}{\bibfnamefont{M.}~\bibnamefont{{Boulay}}}, \bibinfo{author}{\bibfnamefont{S.}~\bibnamefont{{Cebri{\'a}n}}}, \bibinfo{author}{\bibfnamefont{L.}~\bibnamefont{{Covi}}}, \bibinfo{author}{\bibfnamefont{G.}~\bibnamefont{{Fiorillo}}}, \bibinfo{author}{\bibfnamefont{A.}~\bibnamefont{{Green}}}, \bibinfo{author}{\bibfnamefont{J.}~\bibnamefont{{Kopp}}}, \bibinfo{author}{\bibfnamefont{B.}~\bibnamefont{{Majorovits}}}, \bibinfo{author}{\bibfnamefont{K.}~\bibnamefont{{Palladino}}}, \bibinfo{author}{\bibfnamefont{F.}~\bibnamefont{{Petricca}}}, \bibnamefont{et~al.}, \bibinfo{journal}{Reports on Progress in Physics} \textbf{\bibinfo{volume}{85}}, \bibinfo{eid}{056201} (\bibinfo{year}{2022}), \eprint{2104.07634}.

\bibitem[{\citenamefont{{Del Popolo}}(2014)}]{2014IJMPD..2330005D}
\bibinfo{author}{\bibfnamefont{A.}~\bibnamefont{{Del Popolo}}}, \bibinfo{journal}{International Journal of Modern Physics D} \textbf{\bibinfo{volume}{23}}, \bibinfo{eid}{1430005} (\bibinfo{year}{2014}), \eprint{1305.0456}.

\bibitem[{\citenamefont{{Young}}(2017)}]{2017FrPhy..12l1201Y}
\bibinfo{author}{\bibfnamefont{B.-L.} \bibnamefont{{Young}}}, \bibinfo{journal}{Frontiers of Physics} \textbf{\bibinfo{volume}{12}}, \bibinfo{eid}{121201} (\bibinfo{year}{2017}).

\bibitem[{\citenamefont{{Feng}}(2010)}]{2010ARA&A..48..495F}
\bibinfo{author}{\bibfnamefont{J.~L.} \bibnamefont{{Feng}}}, \bibinfo{journal}{\araa} \textbf{\bibinfo{volume}{48}}, \bibinfo{pages}{495} (\bibinfo{year}{2010}), \eprint{1003.0904}.

\bibitem[{\citenamefont{{Baer} et~al.}(2015)\citenamefont{{Baer}, {Choi}, {Kim}, and {Roszkowski}}}]{2015PhR...555....1B}
\bibinfo{author}{\bibfnamefont{H.}~\bibnamefont{{Baer}}}, \bibinfo{author}{\bibfnamefont{K.-Y.} \bibnamefont{{Choi}}}, \bibinfo{author}{\bibfnamefont{J.~E.} \bibnamefont{{Kim}}}, \bibnamefont{and} \bibinfo{author}{\bibfnamefont{L.}~\bibnamefont{{Roszkowski}}}, \bibinfo{journal}{\physrep} \textbf{\bibinfo{volume}{555}}, \bibinfo{pages}{1} (\bibinfo{year}{2015}), \eprint{1407.0017}.

\bibitem[{\citenamefont{{Roszkowski} et~al.}(2018)\citenamefont{{Roszkowski}, {Sessolo}, and {Trojanowski}}}]{2018RPPh...81f6201R}
\bibinfo{author}{\bibfnamefont{L.}~\bibnamefont{{Roszkowski}}}, \bibinfo{author}{\bibfnamefont{E.~M.} \bibnamefont{{Sessolo}}}, \bibnamefont{and} \bibinfo{author}{\bibfnamefont{S.}~\bibnamefont{{Trojanowski}}}, \bibinfo{journal}{Reports on Progress in Physics} \textbf{\bibinfo{volume}{81}}, \bibinfo{eid}{066201} (\bibinfo{year}{2018}), \eprint{1707.06277}.

\bibitem[{\citenamefont{{Profumo} et~al.}(2019)\citenamefont{{Profumo}, {Giani}, and {Piattella}}}]{2019Univ....5..213P}
\bibinfo{author}{\bibfnamefont{S.}~\bibnamefont{{Profumo}}}, \bibinfo{author}{\bibfnamefont{L.}~\bibnamefont{{Giani}}}, \bibnamefont{and} \bibinfo{author}{\bibfnamefont{O.~F.} \bibnamefont{{Piattella}}}, \bibinfo{journal}{Universe} \textbf{\bibinfo{volume}{5}}, \bibinfo{pages}{213} (\bibinfo{year}{2019}), \eprint{1910.05610}.

\bibitem[{\citenamefont{{Kouvaris}}(2008)}]{2008PhRvD..77b3006K}
\bibinfo{author}{\bibfnamefont{C.}~\bibnamefont{{Kouvaris}}}, \bibinfo{journal}{\prd} \textbf{\bibinfo{volume}{77}}, \bibinfo{eid}{023006} (\bibinfo{year}{2008}), \eprint{0708.2362}.

\bibitem[{\citenamefont{{Kouvaris} and {Tinyakov}}(2010)}]{2010PhRvD..82f3531K}
\bibinfo{author}{\bibfnamefont{C.}~\bibnamefont{{Kouvaris}}} \bibnamefont{and} \bibinfo{author}{\bibfnamefont{P.}~\bibnamefont{{Tinyakov}}}, \bibinfo{journal}{\prd} \textbf{\bibinfo{volume}{82}}, \bibinfo{eid}{063531} (\bibinfo{year}{2010}), \eprint{1004.0586}.

\bibitem[{\citenamefont{{Panotopoulos} and {Lopes}}(2017{\natexlab{a}})}]{2017PhRvD..96h3004P}
\bibinfo{author}{\bibfnamefont{G.}~\bibnamefont{{Panotopoulos}}} \bibnamefont{and} \bibinfo{author}{\bibfnamefont{I.}~\bibnamefont{{Lopes}}}, \bibinfo{journal}{\prd} \textbf{\bibinfo{volume}{96}}, \bibinfo{eid}{083004} (\bibinfo{year}{2017}{\natexlab{a}}), \eprint{1709.06312}.

\bibitem[{\citenamefont{{Ivanytskyi} et~al.}(2020)\citenamefont{{Ivanytskyi}, {Sagun}, and {Lopes}}}]{2020PhRvD.102f3028I}
\bibinfo{author}{\bibfnamefont{O.}~\bibnamefont{{Ivanytskyi}}}, \bibinfo{author}{\bibfnamefont{V.}~\bibnamefont{{Sagun}}}, \bibnamefont{and} \bibinfo{author}{\bibfnamefont{I.}~\bibnamefont{{Lopes}}}, \bibinfo{journal}{\prd} \textbf{\bibinfo{volume}{102}}, \bibinfo{eid}{063028} (\bibinfo{year}{2020}), \eprint{1910.09925}.

\bibitem[{\citenamefont{Xiang et~al.}(2014)\citenamefont{Xiang, Jiang, Zhang, and Yang}}]{PhysRevC.89.025803}
\bibinfo{author}{\bibfnamefont{Q.-F.} \bibnamefont{Xiang}}, \bibinfo{author}{\bibfnamefont{W.-Z.} \bibnamefont{Jiang}}, \bibinfo{author}{\bibfnamefont{D.-R.} \bibnamefont{Zhang}}, \bibnamefont{and} \bibinfo{author}{\bibfnamefont{R.-Y.} \bibnamefont{Yang}}, \bibinfo{journal}{Phys. Rev. C} \textbf{\bibinfo{volume}{89}}, \bibinfo{pages}{025803} (\bibinfo{year}{2014}), \urlprefix\url{https://link.aps.org/doi/10.1103/PhysRevC.89.025803}.

\bibitem[{\citenamefont{Das et~al.}(2019)\citenamefont{Das, Malik, and Nayak}}]{PhysRevD.99.043016}
\bibinfo{author}{\bibfnamefont{A.}~\bibnamefont{Das}}, \bibinfo{author}{\bibfnamefont{T.}~\bibnamefont{Malik}}, \bibnamefont{and} \bibinfo{author}{\bibfnamefont{A.~C.} \bibnamefont{Nayak}}, \bibinfo{journal}{Phys. Rev. D} \textbf{\bibinfo{volume}{99}}, \bibinfo{pages}{043016} (\bibinfo{year}{2019}), \urlprefix\url{https://link.aps.org/doi/10.1103/PhysRevD.99.043016}.

\bibitem[{\citenamefont{{Quddus} et~al.}(2020)\citenamefont{{Quddus}, {Panotopoulos}, {Kumar}, {Ahmad}, and {Patra}}}]{2020JPhG...47i5202Q}
\bibinfo{author}{\bibfnamefont{A.}~\bibnamefont{{Quddus}}}, \bibinfo{author}{\bibfnamefont{G.}~\bibnamefont{{Panotopoulos}}}, \bibinfo{author}{\bibfnamefont{B.}~\bibnamefont{{Kumar}}}, \bibinfo{author}{\bibfnamefont{S.}~\bibnamefont{{Ahmad}}}, \bibnamefont{and} \bibinfo{author}{\bibfnamefont{S.~K.} \bibnamefont{{Patra}}}, \bibinfo{journal}{Journal of Physics G Nuclear Physics} \textbf{\bibinfo{volume}{47}}, \bibinfo{eid}{095202} (\bibinfo{year}{2020}), \eprint{1902.00929}.

\bibitem[{\citenamefont{Das et~al.}(2021{\natexlab{a}})\citenamefont{Das, Kumar, and Patra}}]{PhysRevD.104.063028}
\bibinfo{author}{\bibfnamefont{H.~C.} \bibnamefont{Das}}, \bibinfo{author}{\bibfnamefont{A.}~\bibnamefont{Kumar}}, \bibnamefont{and} \bibinfo{author}{\bibfnamefont{S.~K.} \bibnamefont{Patra}}, \bibinfo{journal}{Phys. Rev. D} \textbf{\bibinfo{volume}{104}}, \bibinfo{pages}{063028} (\bibinfo{year}{2021}{\natexlab{a}}), \urlprefix\url{https://link.aps.org/doi/10.1103/PhysRevD.104.063028}.

\bibitem[{\citenamefont{Das et~al.}(2021{\natexlab{b}})\citenamefont{Das, Kumar, Biswal, and Patra}}]{PhysRevD.104.123006}
\bibinfo{author}{\bibfnamefont{H.~C.} \bibnamefont{Das}}, \bibinfo{author}{\bibfnamefont{A.}~\bibnamefont{Kumar}}, \bibinfo{author}{\bibfnamefont{S.~K.} \bibnamefont{Biswal}}, \bibnamefont{and} \bibinfo{author}{\bibfnamefont{S.~K.} \bibnamefont{Patra}}, \bibinfo{journal}{Phys. Rev. D} \textbf{\bibinfo{volume}{104}}, \bibinfo{pages}{123006} (\bibinfo{year}{2021}{\natexlab{b}}), \urlprefix\url{https://link.aps.org/doi/10.1103/PhysRevD.104.123006}.

\bibitem[{\citenamefont{Louren\ifmmode~\mbox{\c{c}}\else \c{c}\fi{}o et~al.}(2022{\natexlab{a}})\citenamefont{Louren\ifmmode~\mbox{\c{c}}\else \c{c}\fi{}o, Frederico, and Dutra}}]{PhysRevD.105.023008}
\bibinfo{author}{\bibfnamefont{O.}~\bibnamefont{Louren\ifmmode~\mbox{\c{c}}\else \c{c}\fi{}o}}, \bibinfo{author}{\bibfnamefont{T.}~\bibnamefont{Frederico}}, \bibnamefont{and} \bibinfo{author}{\bibfnamefont{M.}~\bibnamefont{Dutra}}, \bibinfo{journal}{Phys. Rev. D} \textbf{\bibinfo{volume}{105}}, \bibinfo{pages}{023008} (\bibinfo{year}{2022}{\natexlab{a}}), \urlprefix\url{https://link.aps.org/doi/10.1103/PhysRevD.105.023008}.

\bibitem[{\citenamefont{Dengler et~al.}(2022)\citenamefont{Dengler, Schaffner-Bielich, and Tolos}}]{PhysRevD.105.043013}
\bibinfo{author}{\bibfnamefont{Y.}~\bibnamefont{Dengler}}, \bibinfo{author}{\bibfnamefont{J.}~\bibnamefont{Schaffner-Bielich}}, \bibnamefont{and} \bibinfo{author}{\bibfnamefont{L.}~\bibnamefont{Tolos}}, \bibinfo{journal}{Phys. Rev. D} \textbf{\bibinfo{volume}{105}}, \bibinfo{pages}{043013} (\bibinfo{year}{2022}), \urlprefix\url{https://link.aps.org/doi/10.1103/PhysRevD.105.043013}.

\bibitem[{\citenamefont{Louren\ifmmode~\mbox{\c{c}}\else \c{c}\fi{}o et~al.}(2022{\natexlab{b}})\citenamefont{Louren\ifmmode~\mbox{\c{c}}\else \c{c}\fi{}o, Lenzi, Frederico, and Dutra}}]{PhysRevD.106.043010}
\bibinfo{author}{\bibfnamefont{O.}~\bibnamefont{Louren\ifmmode~\mbox{\c{c}}\else \c{c}\fi{}o}}, \bibinfo{author}{\bibfnamefont{C.~H.} \bibnamefont{Lenzi}}, \bibinfo{author}{\bibfnamefont{T.}~\bibnamefont{Frederico}}, \bibnamefont{and} \bibinfo{author}{\bibfnamefont{M.}~\bibnamefont{Dutra}}, \bibinfo{journal}{Phys. Rev. D} \textbf{\bibinfo{volume}{106}}, \bibinfo{pages}{043010} (\bibinfo{year}{2022}{\natexlab{b}}), \urlprefix\url{https://link.aps.org/doi/10.1103/PhysRevD.106.043010}.

\bibitem[{\citenamefont{{Das} et~al.}(2022)\citenamefont{{Das}, {Malik}, and {Nayak}}}]{2022PhRvD.105l3034D}
\bibinfo{author}{\bibfnamefont{A.}~\bibnamefont{{Das}}}, \bibinfo{author}{\bibfnamefont{T.}~\bibnamefont{{Malik}}}, \bibnamefont{and} \bibinfo{author}{\bibfnamefont{A.~C.} \bibnamefont{{Nayak}}}, \bibinfo{journal}{\prd} \textbf{\bibinfo{volume}{105}}, \bibinfo{eid}{123034} (\bibinfo{year}{2022}).

\bibitem[{\citenamefont{{Panotopoulos} and {Lopes}}(2018)}]{2018IJMPD..2750093P}
\bibinfo{author}{\bibfnamefont{G.}~\bibnamefont{{Panotopoulos}}} \bibnamefont{and} \bibinfo{author}{\bibfnamefont{I.}~\bibnamefont{{Lopes}}}, \bibinfo{journal}{International Journal of Modern Physics D} \textbf{\bibinfo{volume}{27}}, \bibinfo{eid}{1850093} (\bibinfo{year}{2018}), \eprint{1804.05023}.

\bibitem[{\citenamefont{Rafiei~Karkevandi et~al.}(2022)\citenamefont{Rafiei~Karkevandi, Shakeri, Sagun, and Ivanytskyi}}]{PhysRevD.105.023001}
\bibinfo{author}{\bibfnamefont{D.}~\bibnamefont{Rafiei~Karkevandi}}, \bibinfo{author}{\bibfnamefont{S.}~\bibnamefont{Shakeri}}, \bibinfo{author}{\bibfnamefont{V.}~\bibnamefont{Sagun}}, \bibnamefont{and} \bibinfo{author}{\bibfnamefont{O.}~\bibnamefont{Ivanytskyi}}, \bibinfo{journal}{Phys. Rev. D} \textbf{\bibinfo{volume}{105}}, \bibinfo{pages}{023001} (\bibinfo{year}{2022}), \urlprefix\url{https://link.aps.org/doi/10.1103/PhysRevD.105.023001}.

\bibitem[{\citenamefont{Ellis et~al.}(2018)\citenamefont{Ellis, H\"utsi, Kannike, Marzola, Raidal, and Vaskonen}}]{PhysRevD.97.123007}
\bibinfo{author}{\bibfnamefont{J.}~\bibnamefont{Ellis}}, \bibinfo{author}{\bibfnamefont{G.}~\bibnamefont{H\"utsi}}, \bibinfo{author}{\bibfnamefont{K.}~\bibnamefont{Kannike}}, \bibinfo{author}{\bibfnamefont{L.}~\bibnamefont{Marzola}}, \bibinfo{author}{\bibfnamefont{M.}~\bibnamefont{Raidal}}, \bibnamefont{and} \bibinfo{author}{\bibfnamefont{V.}~\bibnamefont{Vaskonen}}, \bibinfo{journal}{Phys. Rev. D} \textbf{\bibinfo{volume}{97}}, \bibinfo{pages}{123007} (\bibinfo{year}{2018}), \urlprefix\url{https://link.aps.org/doi/10.1103/PhysRevD.97.123007}.

\bibitem[{\citenamefont{{Lattimer}}(2021)}]{2021ARNPS..71..433L}
\bibinfo{author}{\bibfnamefont{J.~M.} \bibnamefont{{Lattimer}}}, \bibinfo{journal}{Annual Review of Nuclear and Particle Science} \textbf{\bibinfo{volume}{71}}, \bibinfo{pages}{433} (\bibinfo{year}{2021}).

\bibitem[{\citenamefont{{Del Popolo} et~al.}(2020)\citenamefont{{Del Popolo}, {Le Delliou}, and {Deliyergiyev}}}]{2020Univ....6..222D}
\bibinfo{author}{\bibfnamefont{A.}~\bibnamefont{{Del Popolo}}}, \bibinfo{author}{\bibfnamefont{M.}~\bibnamefont{{Le Delliou}}}, \bibnamefont{and} \bibinfo{author}{\bibfnamefont{M.}~\bibnamefont{{Deliyergiyev}}}, \bibinfo{journal}{Universe} \textbf{\bibinfo{volume}{6}}, \bibinfo{pages}{222} (\bibinfo{year}{2020}).

\bibitem[{\citenamefont{{Herrera} and {Santos}}(1997)}]{1997PhR...286...53H}
\bibinfo{author}{\bibfnamefont{L.}~\bibnamefont{{Herrera}}} \bibnamefont{and} \bibinfo{author}{\bibfnamefont{N.~O.} \bibnamefont{{Santos}}}, \bibinfo{journal}{\physrep} \textbf{\bibinfo{volume}{286}}, \bibinfo{pages}{53} (\bibinfo{year}{1997}).

\bibitem[{\citenamefont{{Kumar} and {Bharti}}(2022)}]{2022NewAR..9501662K}
\bibinfo{author}{\bibfnamefont{J.}~\bibnamefont{{Kumar}}} \bibnamefont{and} \bibinfo{author}{\bibfnamefont{P.}~\bibnamefont{{Bharti}}}, \bibinfo{journal}{\nar} \textbf{\bibinfo{volume}{95}}, \bibinfo{eid}{101662} (\bibinfo{year}{2022}).

\bibitem[{\citenamefont{{Grammenos} et~al.}(2016)\citenamefont{{Grammenos}, {Rahamn}, {Ray}, {Deb}, and {Chowdhury}}}]{2016arXiv161102253G}
\bibinfo{author}{\bibfnamefont{T.}~\bibnamefont{{Grammenos}}}, \bibinfo{author}{\bibfnamefont{F.}~\bibnamefont{{Rahamn}}}, \bibinfo{author}{\bibfnamefont{S.}~\bibnamefont{{Ray}}}, \bibinfo{author}{\bibfnamefont{D.}~\bibnamefont{{Deb}}}, \bibnamefont{and} \bibinfo{author}{\bibfnamefont{S.~R.} \bibnamefont{{Chowdhury}}}, \bibinfo{journal}{arXiv e-prints} \bibinfo{eid}{arXiv:1611.02253} (\bibinfo{year}{2016}), \eprint{1611.02253}.

\bibitem[{\citenamefont{{Rahmansyah} and {Sulaksono}}(2021)}]{2021JPhCS1816a2025R}
\bibinfo{author}{\bibfnamefont{A.}~\bibnamefont{{Rahmansyah}}} \bibnamefont{and} \bibinfo{author}{\bibfnamefont{A.}~\bibnamefont{{Sulaksono}}}, in \emph{\bibinfo{booktitle}{Journal of Physics Conference Series}} (\bibinfo{year}{2021}), vol. \bibinfo{volume}{1816} of \emph{\bibinfo{series}{Journal of Physics Conference Series}}, p. \bibinfo{pages}{012025}.

\bibitem[{\citenamefont{{Moraes} et~al.}(2021)\citenamefont{{Moraes}, {Panotopoulos}, and {Lopes}}}]{2021PhRvD.103h4023M}
\bibinfo{author}{\bibfnamefont{P.~H.~R.~S.} \bibnamefont{{Moraes}}}, \bibinfo{author}{\bibfnamefont{G.}~\bibnamefont{{Panotopoulos}}}, \bibnamefont{and} \bibinfo{author}{\bibfnamefont{I.}~\bibnamefont{{Lopes}}}, \bibinfo{journal}{\prd} \textbf{\bibinfo{volume}{103}}, \bibinfo{eid}{084023} (\bibinfo{year}{2021}), \eprint{2101.02207}.

\bibitem[{\citenamefont{{Michie}}(1963)}]{1963MNRAS.125..127M}
\bibinfo{author}{\bibfnamefont{R.~W.} \bibnamefont{{Michie}}}, \bibinfo{journal}{\mnras} \textbf{\bibinfo{volume}{125}}, \bibinfo{pages}{127} (\bibinfo{year}{1963}).

\bibitem[{\citenamefont{{Binney} and {Tremaine}}(1987)}]{1987gady.book.....B}
\bibinfo{author}{\bibfnamefont{J.}~\bibnamefont{{Binney}}} \bibnamefont{and} \bibinfo{author}{\bibfnamefont{S.}~\bibnamefont{{Tremaine}}}, \emph{\bibinfo{title}{{Galactic dynamics}}} (\bibinfo{year}{1987}).

\bibitem[{\citenamefont{{Cuddeford}}(1991)}]{1991MNRAS.253..414C}
\bibinfo{author}{\bibfnamefont{P.}~\bibnamefont{{Cuddeford}}}, \bibinfo{journal}{\mnras} \textbf{\bibinfo{volume}{253}}, \bibinfo{pages}{414} (\bibinfo{year}{1991}).

\bibitem[{\citenamefont{{Herrera} et~al.}(2002)\citenamefont{{Herrera}, {Martin}, and {Ospino}}}]{2002JMP....43.4889H}
\bibinfo{author}{\bibfnamefont{L.}~\bibnamefont{{Herrera}}}, \bibinfo{author}{\bibfnamefont{J.}~\bibnamefont{{Martin}}}, \bibnamefont{and} \bibinfo{author}{\bibfnamefont{J.}~\bibnamefont{{Ospino}}}, \bibinfo{journal}{Journal of Mathematical Physics} \textbf{\bibinfo{volume}{43}}, \bibinfo{pages}{4889} (\bibinfo{year}{2002}), \eprint{gr-qc/0207040}.

\bibitem[{\citenamefont{{Herrera} et~al.}(2004)\citenamefont{{Herrera}, {di Prisco}, {Martin}, {Ospino}, {Santos}, and {Troconis}}}]{2004PhRvD..69h4026H}
\bibinfo{author}{\bibfnamefont{L.}~\bibnamefont{{Herrera}}}, \bibinfo{author}{\bibfnamefont{A.}~\bibnamefont{{di Prisco}}}, \bibinfo{author}{\bibfnamefont{J.}~\bibnamefont{{Martin}}}, \bibinfo{author}{\bibfnamefont{J.}~\bibnamefont{{Ospino}}}, \bibinfo{author}{\bibfnamefont{N.~O.} \bibnamefont{{Santos}}}, \bibnamefont{and} \bibinfo{author}{\bibfnamefont{O.}~\bibnamefont{{Troconis}}}, \bibinfo{journal}{\prd} \textbf{\bibinfo{volume}{69}}, \bibinfo{eid}{084026} (\bibinfo{year}{2004}), \eprint{gr-qc/0403006}.

\bibitem[{\citenamefont{{Bayin}}(1982)}]{1982PhRvD..26.1262B}
\bibinfo{author}{\bibfnamefont{S.~S.} \bibnamefont{{Bayin}}}, \bibinfo{journal}{\prd} \textbf{\bibinfo{volume}{26}}, \bibinfo{pages}{1262} (\bibinfo{year}{1982}).

\bibitem[{\citenamefont{{Herrera} et~al.}(2008)\citenamefont{{Herrera}, {Ospino}, and {di Prisco}}}]{2008PhRvD..77b7502H}
\bibinfo{author}{\bibfnamefont{L.}~\bibnamefont{{Herrera}}}, \bibinfo{author}{\bibfnamefont{J.}~\bibnamefont{{Ospino}}}, \bibnamefont{and} \bibinfo{author}{\bibfnamefont{A.}~\bibnamefont{{di Prisco}}}, \bibinfo{journal}{\prd} \textbf{\bibinfo{volume}{77}}, \bibinfo{eid}{027502} (\bibinfo{year}{2008}), \eprint{0712.0713}.

\bibitem[{\citenamefont{{Mak} and {Harko}}(2003)}]{2003RSPSA.459..393M}
\bibinfo{author}{\bibfnamefont{M.~K.} \bibnamefont{{Mak}}} \bibnamefont{and} \bibinfo{author}{\bibfnamefont{T.}~\bibnamefont{{Harko}}}, \bibinfo{journal}{Proceedings of the Royal Society of London Series A} \textbf{\bibinfo{volume}{459}}, \bibinfo{pages}{393} (\bibinfo{year}{2003}), \eprint{gr-qc/0110103}.

\bibitem[{\citenamefont{{Lopes} et~al.}(2019)\citenamefont{{Lopes}, {Panotopoulos}, and {Rinc{\'o}n}}}]{2019EPJP..134..454L}
\bibinfo{author}{\bibfnamefont{I.}~\bibnamefont{{Lopes}}}, \bibinfo{author}{\bibfnamefont{G.}~\bibnamefont{{Panotopoulos}}}, \bibnamefont{and} \bibinfo{author}{\bibfnamefont{{\'A}.}~\bibnamefont{{Rinc{\'o}n}}}, \bibinfo{journal}{European Physical Journal Plus} \textbf{\bibinfo{volume}{134}}, \bibinfo{eid}{454} (\bibinfo{year}{2019}), \eprint{1907.03549}.

\bibitem[{\citenamefont{{Sharma} and {Maharaj}}(2007)}]{2007MNRAS.375.1265S}
\bibinfo{author}{\bibfnamefont{R.}~\bibnamefont{{Sharma}}} \bibnamefont{and} \bibinfo{author}{\bibfnamefont{S.~D.} \bibnamefont{{Maharaj}}}, \bibinfo{journal}{\mnras} \textbf{\bibinfo{volume}{375}}, \bibinfo{pages}{1265} (\bibinfo{year}{2007}), \eprint{gr-qc/0702046}.

\bibitem[{\citenamefont{{Oppenheimer} and {Volkoff}}(1939)}]{1939PhRv...55..374O}
\bibinfo{author}{\bibfnamefont{J.~R.} \bibnamefont{{Oppenheimer}}} \bibnamefont{and} \bibinfo{author}{\bibfnamefont{G.~M.} \bibnamefont{{Volkoff}}}, \bibinfo{journal}{Physical Review} \textbf{\bibinfo{volume}{55}}, \bibinfo{pages}{374} (\bibinfo{year}{1939}).

\bibitem[{\citenamefont{{Tolman}}(1939)}]{1939PhRv...55..364T}
\bibinfo{author}{\bibfnamefont{R.~C.} \bibnamefont{{Tolman}}}, \bibinfo{journal}{Physical Review} \textbf{\bibinfo{volume}{55}}, \bibinfo{pages}{364} (\bibinfo{year}{1939}).

\bibitem[{\citenamefont{{Panotopoulos} and {Lopes}}(2017{\natexlab{b}})}]{2017PhRvD..96b3002P}
\bibinfo{author}{\bibfnamefont{G.}~\bibnamefont{{Panotopoulos}}} \bibnamefont{and} \bibinfo{author}{\bibfnamefont{I.}~\bibnamefont{{Lopes}}}, \bibinfo{journal}{\prd} \textbf{\bibinfo{volume}{96}}, \bibinfo{eid}{023002} (\bibinfo{year}{2017}{\natexlab{b}}), \eprint{1706.07272}.

\bibitem[{\citenamefont{{Ciarcelluti} and {Sandin}}(2011)}]{2011PhLB..695...19C}
\bibinfo{author}{\bibfnamefont{P.}~\bibnamefont{{Ciarcelluti}}} \bibnamefont{and} \bibinfo{author}{\bibfnamefont{F.}~\bibnamefont{{Sandin}}}, \bibinfo{journal}{Physics Letters B} \textbf{\bibinfo{volume}{695}}, \bibinfo{pages}{19} (\bibinfo{year}{2011}), \eprint{1005.0857}.

\bibitem[{\citenamefont{{Sandin} and {Ciarcelluti}}(2009)}]{2009APh....32..278S}
\bibinfo{author}{\bibfnamefont{F.}~\bibnamefont{{Sandin}}} \bibnamefont{and} \bibinfo{author}{\bibfnamefont{P.}~\bibnamefont{{Ciarcelluti}}}, \bibinfo{journal}{Astroparticle Physics} \textbf{\bibinfo{volume}{32}}, \bibinfo{pages}{278} (\bibinfo{year}{2009}), \eprint{0809.2942}.

\bibitem[{\citenamefont{{Heintzmann} and {Hillebrandt}}(1975)}]{1975A&A....38...51H}
\bibinfo{author}{\bibfnamefont{H.}~\bibnamefont{{Heintzmann}}} \bibnamefont{and} \bibinfo{author}{\bibfnamefont{W.}~\bibnamefont{{Hillebrandt}}}, \bibinfo{journal}{\aap} \textbf{\bibinfo{volume}{38}}, \bibinfo{pages}{51} (\bibinfo{year}{1975}).

\bibitem[{\citenamefont{{Sagun} et~al.}(2019{\natexlab{a}})\citenamefont{{Sagun}, {Lopes}, and {Ivanytskyi}}}]{2019NuPhA.982..883S}
\bibinfo{author}{\bibfnamefont{V.}~\bibnamefont{{Sagun}}}, \bibinfo{author}{\bibfnamefont{I.}~\bibnamefont{{Lopes}}}, \bibnamefont{and} \bibinfo{author}{\bibfnamefont{A.}~\bibnamefont{{Ivanytskyi}}}, \bibinfo{journal}{\nphysa} \textbf{\bibinfo{volume}{982}}, \bibinfo{pages}{883} (\bibinfo{year}{2019}{\natexlab{a}}), \eprint{1810.04245}.

\bibitem[{\citenamefont{{Al-Mamun} et~al.}(2021)\citenamefont{{Al-Mamun}, {Steiner}, {N{\"a}ttil{\"a}}, {Lange}, {O'Shaughnessy}, {Tews}, {Gandolfi}, {Heinke}, and {Han}}}]{2021PhRvL.126f1101A}
\bibinfo{author}{\bibfnamefont{M.}~\bibnamefont{{Al-Mamun}}}, \bibinfo{author}{\bibfnamefont{A.~W.} \bibnamefont{{Steiner}}}, \bibinfo{author}{\bibfnamefont{J.}~\bibnamefont{{N{\"a}ttil{\"a}}}}, \bibinfo{author}{\bibfnamefont{J.}~\bibnamefont{{Lange}}}, \bibinfo{author}{\bibfnamefont{R.}~\bibnamefont{{O'Shaughnessy}}}, \bibinfo{author}{\bibfnamefont{I.}~\bibnamefont{{Tews}}}, \bibinfo{author}{\bibfnamefont{S.}~\bibnamefont{{Gandolfi}}}, \bibinfo{author}{\bibfnamefont{C.}~\bibnamefont{{Heinke}}}, \bibnamefont{and} \bibinfo{author}{\bibfnamefont{S.}~\bibnamefont{{Han}}}, \bibinfo{journal}{\prl} \textbf{\bibinfo{volume}{126}}, \bibinfo{eid}{061101} (\bibinfo{year}{2021}), \eprint{2008.12817}.

\bibitem[{\citenamefont{{Raaijmakers} et~al.}(2021)\citenamefont{{Raaijmakers}, {Greif}, {Hebeler}, {Hinderer}, {Nissanke}, {Schwenk}, {Riley}, {Watts}, {Lattimer}, and {Ho}}}]{2021ApJ...918L..29R}
\bibinfo{author}{\bibfnamefont{G.}~\bibnamefont{{Raaijmakers}}}, \bibinfo{author}{\bibfnamefont{S.~K.} \bibnamefont{{Greif}}}, \bibinfo{author}{\bibfnamefont{K.}~\bibnamefont{{Hebeler}}}, \bibinfo{author}{\bibfnamefont{T.}~\bibnamefont{{Hinderer}}}, \bibinfo{author}{\bibfnamefont{S.}~\bibnamefont{{Nissanke}}}, \bibinfo{author}{\bibfnamefont{A.}~\bibnamefont{{Schwenk}}}, \bibinfo{author}{\bibfnamefont{T.~E.} \bibnamefont{{Riley}}}, \bibinfo{author}{\bibfnamefont{A.~L.} \bibnamefont{{Watts}}}, \bibinfo{author}{\bibfnamefont{J.~M.} \bibnamefont{{Lattimer}}}, \bibnamefont{and} \bibinfo{author}{\bibfnamefont{W.~C.~G.} \bibnamefont{{Ho}}}, \bibinfo{journal}{\apjl} \textbf{\bibinfo{volume}{918}}, \bibinfo{eid}{L29} (\bibinfo{year}{2021}), \eprint{2105.06981}.

\bibitem[{\citenamefont{{Pratt} et~al.}(2015)\citenamefont{{Pratt}, {Sangaline}, {Sorensen}, and {Wang}}}]{2015PhRvL.114t2301P}
\bibinfo{author}{\bibfnamefont{S.}~\bibnamefont{{Pratt}}}, \bibinfo{author}{\bibfnamefont{E.}~\bibnamefont{{Sangaline}}}, \bibinfo{author}{\bibfnamefont{P.}~\bibnamefont{{Sorensen}}}, \bibnamefont{and} \bibinfo{author}{\bibfnamefont{H.}~\bibnamefont{{Wang}}}, \bibinfo{journal}{\prl} \textbf{\bibinfo{volume}{114}}, \bibinfo{eid}{202301} (\bibinfo{year}{2015}), \eprint{1501.04042}.

\bibitem[{\citenamefont{{Bazavov} et~al.}(2017)\citenamefont{{Bazavov}, {Ding}, {Hegde}, {Kaczmarek}, {Karsch}, {Laermann}, {Maezawa}, {Mukherjee}, {Ohno}, {Petreczky} et~al.}}]{2017PhRvD..95e4504B}
\bibinfo{author}{\bibfnamefont{A.}~\bibnamefont{{Bazavov}}}, \bibinfo{author}{\bibfnamefont{H.~T.} \bibnamefont{{Ding}}}, \bibinfo{author}{\bibfnamefont{P.}~\bibnamefont{{Hegde}}}, \bibinfo{author}{\bibfnamefont{O.}~\bibnamefont{{Kaczmarek}}}, \bibinfo{author}{\bibfnamefont{F.}~\bibnamefont{{Karsch}}}, \bibinfo{author}{\bibfnamefont{E.}~\bibnamefont{{Laermann}}}, \bibinfo{author}{\bibfnamefont{Y.}~\bibnamefont{{Maezawa}}}, \bibinfo{author}{\bibfnamefont{S.}~\bibnamefont{{Mukherjee}}}, \bibinfo{author}{\bibfnamefont{H.}~\bibnamefont{{Ohno}}}, \bibinfo{author}{\bibfnamefont{P.}~\bibnamefont{{Petreczky}}}, \bibnamefont{et~al.}, \bibinfo{journal}{\prd} \textbf{\bibinfo{volume}{95}}, \bibinfo{eid}{054504} (\bibinfo{year}{2017}), \eprint{1701.04325}.

\bibitem[{\citenamefont{{Haque} et~al.}(2014)\citenamefont{{Haque}, {Bandyopadhyay}, {Andersen}, {Mustafa}, {Strickland}, and {Su}}}]{2014JHEP...05..027H}
\bibinfo{author}{\bibfnamefont{N.}~\bibnamefont{{Haque}}}, \bibinfo{author}{\bibfnamefont{A.}~\bibnamefont{{Bandyopadhyay}}}, \bibinfo{author}{\bibfnamefont{J.~O.} \bibnamefont{{Andersen}}}, \bibinfo{author}{\bibfnamefont{M.~G.} \bibnamefont{{Mustafa}}}, \bibinfo{author}{\bibfnamefont{M.}~\bibnamefont{{Strickland}}}, \bibnamefont{and} \bibinfo{author}{\bibfnamefont{N.}~\bibnamefont{{Su}}}, \bibinfo{journal}{Journal of High Energy Physics} \textbf{\bibinfo{volume}{2014}}, \bibinfo{eid}{27} (\bibinfo{year}{2014}), \eprint{1402.6907}.

\bibitem[{\citenamefont{Papazoglou et~al.}(1999)\citenamefont{Papazoglou, Zschiesche, Schramm, Schaffner-Bielich, Stoecker, and Greiner}}]{Papazoglou:1998vr}
\bibinfo{author}{\bibfnamefont{P.}~\bibnamefont{Papazoglou}}, \bibinfo{author}{\bibfnamefont{D.}~\bibnamefont{Zschiesche}}, \bibinfo{author}{\bibfnamefont{S.}~\bibnamefont{Schramm}}, \bibinfo{author}{\bibfnamefont{J.}~\bibnamefont{Schaffner-Bielich}}, \bibinfo{author}{\bibfnamefont{H.}~\bibnamefont{Stoecker}}, \bibnamefont{and} \bibinfo{author}{\bibfnamefont{W.}~\bibnamefont{Greiner}}, \bibinfo{journal}{Phys. Rev. C} \textbf{\bibinfo{volume}{59}}, \bibinfo{pages}{411} (\bibinfo{year}{1999}), \eprint{nucl-th/9806087}.

\bibitem[{\citenamefont{Dexheimer and Schramm}(2008)}]{Dexheimer:2008ax}
\bibinfo{author}{\bibfnamefont{V.}~\bibnamefont{Dexheimer}} \bibnamefont{and} \bibinfo{author}{\bibfnamefont{S.}~\bibnamefont{Schramm}}, \bibinfo{journal}{Astrophys. J.} \textbf{\bibinfo{volume}{683}}, \bibinfo{pages}{943} (\bibinfo{year}{2008}), \eprint{0802.1999}.

\bibitem[{\citenamefont{Motornenko et~al.}(2020)\citenamefont{Motornenko, Steinheimer, Vovchenko, Schramm, and Stoecker}}]{PhysRevC.101.034904}
\bibinfo{author}{\bibfnamefont{A.}~\bibnamefont{Motornenko}}, \bibinfo{author}{\bibfnamefont{J.}~\bibnamefont{Steinheimer}}, \bibinfo{author}{\bibfnamefont{V.}~\bibnamefont{Vovchenko}}, \bibinfo{author}{\bibfnamefont{S.}~\bibnamefont{Schramm}}, \bibnamefont{and} \bibinfo{author}{\bibfnamefont{H.}~\bibnamefont{Stoecker}}, \bibinfo{journal}{Phys. Rev. C} \textbf{\bibinfo{volume}{101}}, \bibinfo{pages}{034904} (\bibinfo{year}{2020}), \urlprefix\url{https://link.aps.org/doi/10.1103/PhysRevC.101.034904}.

\bibitem[{\citenamefont{Motornenko et~al.}(2021)\citenamefont{Motornenko, Pal, Bhattacharyya, Steinheimer, and Stoecker}}]{Motornenko:2020yme}
\bibinfo{author}{\bibfnamefont{A.}~\bibnamefont{Motornenko}}, \bibinfo{author}{\bibfnamefont{S.}~\bibnamefont{Pal}}, \bibinfo{author}{\bibfnamefont{A.}~\bibnamefont{Bhattacharyya}}, \bibinfo{author}{\bibfnamefont{J.}~\bibnamefont{Steinheimer}}, \bibnamefont{and} \bibinfo{author}{\bibfnamefont{H.}~\bibnamefont{Stoecker}}, \bibinfo{journal}{Phys. Rev. C} \textbf{\bibinfo{volume}{103}}, \bibinfo{pages}{054908} (\bibinfo{year}{2021}), \eprint{2009.10848}.

\bibitem[{\citenamefont{Group et~al.}(2022)\citenamefont{Group, Workman, Burkert, Crede, Klempt, Thoma, Tiator, Agashe, Aielli, Allanach et~al.}}]{PTEP2022}
\bibinfo{author}{\bibfnamefont{P.~D.} \bibnamefont{Group}}, \bibinfo{author}{\bibfnamefont{R.~L.} \bibnamefont{Workman}}, \bibinfo{author}{\bibfnamefont{V.~D.} \bibnamefont{Burkert}}, \bibinfo{author}{\bibfnamefont{V.}~\bibnamefont{Crede}}, \bibinfo{author}{\bibfnamefont{E.}~\bibnamefont{Klempt}}, \bibinfo{author}{\bibfnamefont{U.}~\bibnamefont{Thoma}}, \bibinfo{author}{\bibfnamefont{L.}~\bibnamefont{Tiator}}, \bibinfo{author}{\bibfnamefont{K.}~\bibnamefont{Agashe}}, \bibinfo{author}{\bibfnamefont{G.}~\bibnamefont{Aielli}}, \bibinfo{author}{\bibfnamefont{B.~C.} \bibnamefont{Allanach}}, \bibnamefont{et~al.}, \bibinfo{journal}{Progress of Theoretical and Experimental Physics} \textbf{\bibinfo{volume}{2022}}, \bibinfo{pages}{083C01} (\bibinfo{year}{2022}), ISSN \bibinfo{issn}{2050-3911}, \eprint{https://academic.oup.com/ptep/article-pdf/2022/8/083C01/49175539/ptac097.pdf}, \urlprefix\url{https://doi.org/10.1093/ptep/ptac097}.

\bibitem[{\citenamefont{Zyla et~al.}(2020)}]{ParticleDataGroup:2020ssz}
\bibinfo{author}{\bibfnamefont{P.~A.} \bibnamefont{Zyla}} \bibnamefont{et~al.} (\bibinfo{collaboration}{Particle Data Group}), \bibinfo{journal}{PTEP} \textbf{\bibinfo{volume}{2020}}, \bibinfo{pages}{083C01} (\bibinfo{year}{2020}).

\bibitem[{\citenamefont{Schneider et~al.}(2017)\citenamefont{Schneider, Roberts, and Ott}}]{Schneider:2017tfi}
\bibinfo{author}{\bibfnamefont{A.~S.} \bibnamefont{Schneider}}, \bibinfo{author}{\bibfnamefont{L.~F.} \bibnamefont{Roberts}}, \bibnamefont{and} \bibinfo{author}{\bibfnamefont{C.~D.} \bibnamefont{Ott}}, \bibinfo{journal}{Phys. Rev. C} \textbf{\bibinfo{volume}{96}}, \bibinfo{pages}{065802} (\bibinfo{year}{2017}), \eprint{1707.01527}.

\bibitem[{\citenamefont{Motornenko et~al.}(2019)\citenamefont{Motornenko, Vovchenko, Steinheimer, Schramm, and Stoecker}}]{Motornenko:2018hjw}
\bibinfo{author}{\bibfnamefont{A.}~\bibnamefont{Motornenko}}, \bibinfo{author}{\bibfnamefont{V.}~\bibnamefont{Vovchenko}}, \bibinfo{author}{\bibfnamefont{J.}~\bibnamefont{Steinheimer}}, \bibinfo{author}{\bibfnamefont{S.}~\bibnamefont{Schramm}}, \bibnamefont{and} \bibinfo{author}{\bibfnamefont{H.}~\bibnamefont{Stoecker}}, \bibinfo{journal}{Nucl. Phys. A} \textbf{\bibinfo{volume}{982}}, \bibinfo{pages}{891} (\bibinfo{year}{2019}), \eprint{1809.02000}.

\bibitem[{\citenamefont{Most et~al.}(2023)\citenamefont{Most, Motornenko, Steinheimer, Dexheimer, Hanauske, Rezzolla, and Stoecker}}]{Most:2022wgo}
\bibinfo{author}{\bibfnamefont{E.~R.} \bibnamefont{Most}}, \bibinfo{author}{\bibfnamefont{A.}~\bibnamefont{Motornenko}}, \bibinfo{author}{\bibfnamefont{J.}~\bibnamefont{Steinheimer}}, \bibinfo{author}{\bibfnamefont{V.}~\bibnamefont{Dexheimer}}, \bibinfo{author}{\bibfnamefont{M.}~\bibnamefont{Hanauske}}, \bibinfo{author}{\bibfnamefont{L.}~\bibnamefont{Rezzolla}}, \bibnamefont{and} \bibinfo{author}{\bibfnamefont{H.}~\bibnamefont{Stoecker}}, \bibinfo{journal}{Phys. Rev. D} \textbf{\bibinfo{volume}{107}}, \bibinfo{pages}{043034} (\bibinfo{year}{2023}), \eprint{2201.13150}.

\bibitem[{\citenamefont{Abbott et~al.}(2019)\citenamefont{Abbott, Abbott, Abbott, Acernese, Ackley, Adams, Adams, Addesso, Adhikari, Adya et~al.}}]{PhysRevX.9.011001}
\bibinfo{author}{\bibfnamefont{B.~P.} \bibnamefont{Abbott}}, \bibinfo{author}{\bibfnamefont{R.}~\bibnamefont{Abbott}}, \bibinfo{author}{\bibfnamefont{T.~D.} \bibnamefont{Abbott}}, \bibinfo{author}{\bibfnamefont{F.}~\bibnamefont{Acernese}}, \bibinfo{author}{\bibfnamefont{K.}~\bibnamefont{Ackley}}, \bibinfo{author}{\bibfnamefont{C.}~\bibnamefont{Adams}}, \bibinfo{author}{\bibfnamefont{T.}~\bibnamefont{Adams}}, \bibinfo{author}{\bibfnamefont{P.}~\bibnamefont{Addesso}}, \bibinfo{author}{\bibfnamefont{R.~X.} \bibnamefont{Adhikari}}, \bibinfo{author}{\bibfnamefont{V.~B.} \bibnamefont{Adya}}, \bibnamefont{et~al.} (\bibinfo{collaboration}{LIGO Scientific Collaboration and Virgo Collaboration}), \bibinfo{journal}{Phys. Rev. X} \textbf{\bibinfo{volume}{9}}, \bibinfo{pages}{011001} (\bibinfo{year}{2019}), \urlprefix\url{https://link.aps.org/doi/10.1103/PhysRevX.9.011001}.

\bibitem[{\citenamefont{{Miller} et~al.}(2019)\citenamefont{{Miller}, {Lamb}, {Dittmann}, {Bogdanov}, {Arzoumanian}, {Gendreau}, {Guillot}, {Harding}, {Ho}, {Lattimer} et~al.}}]{2019ApJ...887L..24M}
\bibinfo{author}{\bibfnamefont{M.~C.} \bibnamefont{{Miller}}}, \bibinfo{author}{\bibfnamefont{F.~K.} \bibnamefont{{Lamb}}}, \bibinfo{author}{\bibfnamefont{A.~J.} \bibnamefont{{Dittmann}}}, \bibinfo{author}{\bibfnamefont{S.}~\bibnamefont{{Bogdanov}}}, \bibinfo{author}{\bibfnamefont{Z.}~\bibnamefont{{Arzoumanian}}}, \bibinfo{author}{\bibfnamefont{K.~C.} \bibnamefont{{Gendreau}}}, \bibinfo{author}{\bibfnamefont{S.}~\bibnamefont{{Guillot}}}, \bibinfo{author}{\bibfnamefont{A.~K.} \bibnamefont{{Harding}}}, \bibinfo{author}{\bibfnamefont{W.~C.~G.} \bibnamefont{{Ho}}}, \bibinfo{author}{\bibfnamefont{J.~M.} \bibnamefont{{Lattimer}}}, \bibnamefont{et~al.}, \bibinfo{journal}{\apjl} \textbf{\bibinfo{volume}{887}}, \bibinfo{eid}{L24} (\bibinfo{year}{2019}), \eprint{1912.05705}.

\bibitem[{\citenamefont{{Baym} et~al.}(1971)\citenamefont{{Baym}, {Pethick}, and {Sutherland}}}]{1971ApJ...170..299B}
\bibinfo{author}{\bibfnamefont{G.}~\bibnamefont{{Baym}}}, \bibinfo{author}{\bibfnamefont{C.}~\bibnamefont{{Pethick}}}, \bibnamefont{and} \bibinfo{author}{\bibfnamefont{P.}~\bibnamefont{{Sutherland}}}, \bibinfo{journal}{\apj} \textbf{\bibinfo{volume}{170}}, \bibinfo{pages}{299} (\bibinfo{year}{1971}).

\bibitem[{\citenamefont{{Sagun} et~al.}(2019{\natexlab{b}})\citenamefont{{Sagun}, {Lopes}, and {Ivanytskyi}}}]{2019ApJ...871..157S}
\bibinfo{author}{\bibfnamefont{V.~V.} \bibnamefont{{Sagun}}}, \bibinfo{author}{\bibfnamefont{I.}~\bibnamefont{{Lopes}}}, \bibnamefont{and} \bibinfo{author}{\bibfnamefont{A.~I.} \bibnamefont{{Ivanytskyi}}}, \bibinfo{journal}{\apj} \textbf{\bibinfo{volume}{871}}, \bibinfo{eid}{157} (\bibinfo{year}{2019}{\natexlab{b}}), \eprint{1805.04976}.

\bibitem[{\citenamefont{{Serot} and {Walecka}}(1997)}]{1997IJMPE...6..515S}
\bibinfo{author}{\bibfnamefont{B.~D.} \bibnamefont{{Serot}}} \bibnamefont{and} \bibinfo{author}{\bibfnamefont{J.~D.} \bibnamefont{{Walecka}}}, \bibinfo{journal}{International Journal of Modern Physics E} \textbf{\bibinfo{volume}{6}}, \bibinfo{pages}{515} (\bibinfo{year}{1997}), \eprint{nucl-th/9701058}.

\bibitem[{\citenamefont{{B{\"o}hmer} and {Harko}}(2007)}]{2007JCAP...06..025B}
\bibinfo{author}{\bibfnamefont{C.~G.} \bibnamefont{{B{\"o}hmer}}} \bibnamefont{and} \bibinfo{author}{\bibfnamefont{T.}~\bibnamefont{{Harko}}}, \bibinfo{journal}{\jcap} \textbf{\bibinfo{volume}{2007}}, \bibinfo{eid}{025} (\bibinfo{year}{2007}), \eprint{0705.4158}.

\bibitem[{\citenamefont{{Li} et~al.}(2012)\citenamefont{{Li}, {Harko}, and {Cheng}}}]{2012JCAP...06..001L}
\bibinfo{author}{\bibfnamefont{X.~Y.} \bibnamefont{{Li}}}, \bibinfo{author}{\bibfnamefont{T.}~\bibnamefont{{Harko}}}, \bibnamefont{and} \bibinfo{author}{\bibfnamefont{K.~S.} \bibnamefont{{Cheng}}}, \bibinfo{journal}{\jcap} \textbf{\bibinfo{volume}{2012}}, \bibinfo{eid}{001} (\bibinfo{year}{2012}), \eprint{1205.2932}.

\bibitem[{\citenamefont{{Harko} and {Lobo}}(2015)}]{2015PhRvD..92d3011H}
\bibinfo{author}{\bibfnamefont{T.}~\bibnamefont{{Harko}}} \bibnamefont{and} \bibinfo{author}{\bibfnamefont{F.~S.~N.} \bibnamefont{{Lobo}}}, \bibinfo{journal}{\prd} \textbf{\bibinfo{volume}{92}}, \bibinfo{eid}{043011} (\bibinfo{year}{2015}), \eprint{1505.00944}.

\bibitem[{\citenamefont{{Cr{\v{a}}ciun} and {Harko}}(2020)}]{2020EPJC...80..735C}
\bibinfo{author}{\bibfnamefont{M.}~\bibnamefont{{Cr{\v{a}}ciun}}} \bibnamefont{and} \bibinfo{author}{\bibfnamefont{T.}~\bibnamefont{{Harko}}}, \bibinfo{journal}{European Physical Journal C} \textbf{\bibinfo{volume}{80}}, \bibinfo{eid}{735} (\bibinfo{year}{2020}), \eprint{2007.12222}.

\bibitem[{\citenamefont{{Schunck} and {Mielke}}(2003)}]{2003CQGra..20R.301S}
\bibinfo{author}{\bibfnamefont{F.~E.} \bibnamefont{{Schunck}}} \bibnamefont{and} \bibinfo{author}{\bibfnamefont{E.~W.} \bibnamefont{{Mielke}}}, \bibinfo{journal}{Classical and Quantum Gravity} \textbf{\bibinfo{volume}{20}}, \bibinfo{pages}{R301} (\bibinfo{year}{2003}), \eprint{0801.0307}.

\bibitem[{\citenamefont{{Gleiser} and {Dev}}(2004)}]{2004IJMPD..13.1389G}
\bibinfo{author}{\bibfnamefont{M.}~\bibnamefont{{Gleiser}}} \bibnamefont{and} \bibinfo{author}{\bibfnamefont{K.}~\bibnamefont{{Dev}}}, \bibinfo{journal}{International Journal of Modern Physics D} \textbf{\bibinfo{volume}{13}}, \bibinfo{pages}{1389} (\bibinfo{year}{2004}), \eprint{astro-ph/0401546}.

\bibitem[{\citenamefont{{Barreto}}(1993)}]{1993Ap&SS.201..191B}
\bibinfo{author}{\bibfnamefont{W.}~\bibnamefont{{Barreto}}}, \bibinfo{journal}{\apss} \textbf{\bibinfo{volume}{201}}, \bibinfo{pages}{191} (\bibinfo{year}{1993}).

\bibitem[{\citenamefont{{Letelier}}(1980)}]{1980PhRvD..22..807L}
\bibinfo{author}{\bibfnamefont{P.~S.} \bibnamefont{{Letelier}}}, \bibinfo{journal}{\prd} \textbf{\bibinfo{volume}{22}}, \bibinfo{pages}{807} (\bibinfo{year}{1980}).

\bibitem[{\citenamefont{{Naidu} et~al.}(2022)\citenamefont{{Naidu}, {Carloni}, and {Dunsby}}}]{2022PhRvD.106l4023N}
\bibinfo{author}{\bibfnamefont{N.~F.} \bibnamefont{{Naidu}}}, \bibinfo{author}{\bibfnamefont{S.}~\bibnamefont{{Carloni}}}, \bibnamefont{and} \bibinfo{author}{\bibfnamefont{P.}~\bibnamefont{{Dunsby}}}, \bibinfo{journal}{\prd} \textbf{\bibinfo{volume}{106}}, \bibinfo{eid}{124023} (\bibinfo{year}{2022}), \eprint{2210.06867}.

\bibitem[{\citenamefont{{Le{\'o}n} et~al.}(2023)\citenamefont{{Le{\'o}n}, {Fuenmayor}, and {Contreras}}}]{2023PhRvD.107f4010L}
\bibinfo{author}{\bibfnamefont{P.}~\bibnamefont{{Le{\'o}n}}}, \bibinfo{author}{\bibfnamefont{E.}~\bibnamefont{{Fuenmayor}}}, \bibnamefont{and} \bibinfo{author}{\bibfnamefont{E.}~\bibnamefont{{Contreras}}}, \bibinfo{journal}{\prd} \textbf{\bibinfo{volume}{107}}, \bibinfo{eid}{064010} (\bibinfo{year}{2023}), \eprint{2303.04211}.

\bibitem[{\citenamefont{{Ratanpal} and {Patel}}(2023)}]{2023Ap&SS.368...21R}
\bibinfo{author}{\bibfnamefont{B.~S.} \bibnamefont{{Ratanpal}}} \bibnamefont{and} \bibinfo{author}{\bibfnamefont{R.}~\bibnamefont{{Patel}}}, \bibinfo{journal}{\apss} \textbf{\bibinfo{volume}{368}}, \bibinfo{eid}{21} (\bibinfo{year}{2023}).

\bibitem[{\citenamefont{{Horvat} et~al.}(2011)\citenamefont{{Horvat}, {Iliji{\'c}}, and {Marunovi{\'c}}}}]{2011CQGra..28b5009H}
\bibinfo{author}{\bibfnamefont{D.}~\bibnamefont{{Horvat}}}, \bibinfo{author}{\bibfnamefont{S.}~\bibnamefont{{Iliji{\'c}}}}, \bibnamefont{and} \bibinfo{author}{\bibfnamefont{A.}~\bibnamefont{{Marunovi{\'c}}}}, \bibinfo{journal}{Classical and Quantum Gravity} \textbf{\bibinfo{volume}{28}}, \bibinfo{eid}{025009} (\bibinfo{year}{2011}), \eprint{1010.0878}.

\bibitem[{\citenamefont{{Bowers} and {Liang}}(1974)}]{1974ApJ...188..657B}
\bibinfo{author}{\bibfnamefont{R.~L.} \bibnamefont{{Bowers}}} \bibnamefont{and} \bibinfo{author}{\bibfnamefont{E.~P.~T.} \bibnamefont{{Liang}}}, \bibinfo{journal}{\apj} \textbf{\bibinfo{volume}{188}}, \bibinfo{pages}{657} (\bibinfo{year}{1974}).

\bibitem[{\citenamefont{{Folomeev} and {Dzhunushaliev}}(2015)}]{2015PhRvD..91d4040F}
\bibinfo{author}{\bibfnamefont{V.}~\bibnamefont{{Folomeev}}} \bibnamefont{and} \bibinfo{author}{\bibfnamefont{V.}~\bibnamefont{{Dzhunushaliev}}}, \bibinfo{journal}{\prd} \textbf{\bibinfo{volume}{91}}, \bibinfo{eid}{044040} (\bibinfo{year}{2015}), \eprint{1501.06275}.

\bibitem[{\citenamefont{{Riley} et~al.}(2021)\citenamefont{{Riley}, {Watts}, {Ray}, {Bogdanov}, {Guillot}, {Morsink}, {Bilous}, {Arzoumanian}, {Choudhury}, {Deneva} et~al.}}]{2021ApJ...918L..27R}
\bibinfo{author}{\bibfnamefont{T.~E.} \bibnamefont{{Riley}}}, \bibinfo{author}{\bibfnamefont{A.~L.} \bibnamefont{{Watts}}}, \bibinfo{author}{\bibfnamefont{P.~S.} \bibnamefont{{Ray}}}, \bibinfo{author}{\bibfnamefont{S.}~\bibnamefont{{Bogdanov}}}, \bibinfo{author}{\bibfnamefont{S.}~\bibnamefont{{Guillot}}}, \bibinfo{author}{\bibfnamefont{S.~M.} \bibnamefont{{Morsink}}}, \bibinfo{author}{\bibfnamefont{A.~V.} \bibnamefont{{Bilous}}}, \bibinfo{author}{\bibfnamefont{Z.}~\bibnamefont{{Arzoumanian}}}, \bibinfo{author}{\bibfnamefont{D.}~\bibnamefont{{Choudhury}}}, \bibinfo{author}{\bibfnamefont{J.~S.} \bibnamefont{{Deneva}}}, \bibnamefont{et~al.}, \bibinfo{journal}{\apjl} \textbf{\bibinfo{volume}{918}}, \bibinfo{eid}{L27} (\bibinfo{year}{2021}), \eprint{2105.06980}.

\bibitem[{\citenamefont{{Miller} et~al.}(2021)\citenamefont{{Miller}, {Lamb}, {Dittmann}, {Bogdanov}, {Arzoumanian}, {Gendreau}, {Guillot}, {Ho}, {Lattimer}, {Loewenstein} et~al.}}]{2021ApJ...918L..28M}
\bibinfo{author}{\bibfnamefont{M.~C.} \bibnamefont{{Miller}}}, \bibinfo{author}{\bibfnamefont{F.~K.} \bibnamefont{{Lamb}}}, \bibinfo{author}{\bibfnamefont{A.~J.} \bibnamefont{{Dittmann}}}, \bibinfo{author}{\bibfnamefont{S.}~\bibnamefont{{Bogdanov}}}, \bibinfo{author}{\bibfnamefont{Z.}~\bibnamefont{{Arzoumanian}}}, \bibinfo{author}{\bibfnamefont{K.~C.} \bibnamefont{{Gendreau}}}, \bibinfo{author}{\bibfnamefont{S.}~\bibnamefont{{Guillot}}}, \bibinfo{author}{\bibfnamefont{W.~C.~G.} \bibnamefont{{Ho}}}, \bibinfo{author}{\bibfnamefont{J.~M.} \bibnamefont{{Lattimer}}}, \bibinfo{author}{\bibfnamefont{M.}~\bibnamefont{{Loewenstein}}}, \bibnamefont{et~al.}, \bibinfo{journal}{\apjl} \textbf{\bibinfo{volume}{918}}, \bibinfo{eid}{L28} (\bibinfo{year}{2021}), \eprint{2105.06979}.

\bibitem[{\citenamefont{{Abbott} et~al.}(2018)\citenamefont{{Abbott}, {Abbott}, {Abbott}, {Acernese}, {Ackley}, {Adams}, {Adams}, {Addesso}, {Adhikari}, {Adya} et~al.}}]{2018PhRvL.121p1101A}
\bibinfo{author}{\bibfnamefont{B.~P.} \bibnamefont{{Abbott}}}, \bibinfo{author}{\bibfnamefont{R.}~\bibnamefont{{Abbott}}}, \bibinfo{author}{\bibfnamefont{T.~D.} \bibnamefont{{Abbott}}}, \bibinfo{author}{\bibfnamefont{F.}~\bibnamefont{{Acernese}}}, \bibinfo{author}{\bibfnamefont{K.}~\bibnamefont{{Ackley}}}, \bibinfo{author}{\bibfnamefont{C.}~\bibnamefont{{Adams}}}, \bibinfo{author}{\bibfnamefont{T.}~\bibnamefont{{Adams}}}, \bibinfo{author}{\bibfnamefont{P.}~\bibnamefont{{Addesso}}}, \bibinfo{author}{\bibfnamefont{R.~X.} \bibnamefont{{Adhikari}}}, \bibinfo{author}{\bibfnamefont{V.~B.} \bibnamefont{{Adya}}}, \bibnamefont{et~al.}, \bibinfo{journal}{\prl} \textbf{\bibinfo{volume}{121}}, \bibinfo{eid}{161101} (\bibinfo{year}{2018}), \eprint{1805.11581}.

\bibitem[{\citenamefont{{Clowe} et~al.}(2004)\citenamefont{{Clowe}, {Gonzalez}, and {Markevitch}}}]{2004ApJ...604..596C}
\bibinfo{author}{\bibfnamefont{D.}~\bibnamefont{{Clowe}}}, \bibinfo{author}{\bibfnamefont{A.}~\bibnamefont{{Gonzalez}}}, \bibnamefont{and} \bibinfo{author}{\bibfnamefont{M.}~\bibnamefont{{Markevitch}}}, \bibinfo{journal}{\apj} \textbf{\bibinfo{volume}{604}}, \bibinfo{pages}{596} (\bibinfo{year}{2004}), \eprint{astro-ph/0312273}.

\bibitem[{\citenamefont{{Brada{\v{c}}} et~al.}(2008)\citenamefont{{Brada{\v{c}}}, {Allen}, {Treu}, {Ebeling}, {Massey}, {Morris}, {von der Linden}, and {Applegate}}}]{2008ApJ...687..959B}
\bibinfo{author}{\bibfnamefont{M.}~\bibnamefont{{Brada{\v{c}}}}}, \bibinfo{author}{\bibfnamefont{S.~W.} \bibnamefont{{Allen}}}, \bibinfo{author}{\bibfnamefont{T.}~\bibnamefont{{Treu}}}, \bibinfo{author}{\bibfnamefont{H.}~\bibnamefont{{Ebeling}}}, \bibinfo{author}{\bibfnamefont{R.}~\bibnamefont{{Massey}}}, \bibinfo{author}{\bibfnamefont{R.~G.} \bibnamefont{{Morris}}}, \bibinfo{author}{\bibfnamefont{A.}~\bibnamefont{{von der Linden}}}, \bibnamefont{and} \bibinfo{author}{\bibfnamefont{D.}~\bibnamefont{{Applegate}}}, \bibinfo{journal}{\apj} \textbf{\bibinfo{volume}{687}}, \bibinfo{pages}{959} (\bibinfo{year}{2008}), \eprint{0806.2320}.

\bibitem[{\citenamefont{{Clowe} et~al.}(2006)\citenamefont{{Clowe}, {Brada{\v{c}}}, {Gonzalez}, {Markevitch}, {Randall}, {Jones}, and {Zaritsky}}}]{2006ApJ...648L.109C}
\bibinfo{author}{\bibfnamefont{D.}~\bibnamefont{{Clowe}}}, \bibinfo{author}{\bibfnamefont{M.}~\bibnamefont{{Brada{\v{c}}}}}, \bibinfo{author}{\bibfnamefont{A.~H.} \bibnamefont{{Gonzalez}}}, \bibinfo{author}{\bibfnamefont{M.}~\bibnamefont{{Markevitch}}}, \bibinfo{author}{\bibfnamefont{S.~W.} \bibnamefont{{Randall}}}, \bibinfo{author}{\bibfnamefont{C.}~\bibnamefont{{Jones}}}, \bibnamefont{and} \bibinfo{author}{\bibfnamefont{D.}~\bibnamefont{{Zaritsky}}}, \bibinfo{journal}{\apjl} \textbf{\bibinfo{volume}{648}}, \bibinfo{pages}{L109} (\bibinfo{year}{2006}), \eprint{astro-ph/0608407}.

\bibitem[{\citenamefont{{Thompson} et~al.}(2015)\citenamefont{{Thompson}, {Dav{\'e}}, and {Nagamine}}}]{2015MNRAS.452.3030T}
\bibinfo{author}{\bibfnamefont{R.}~\bibnamefont{{Thompson}}}, \bibinfo{author}{\bibfnamefont{R.}~\bibnamefont{{Dav{\'e}}}}, \bibnamefont{and} \bibinfo{author}{\bibfnamefont{K.}~\bibnamefont{{Nagamine}}}, \bibinfo{journal}{\mnras} \textbf{\bibinfo{volume}{452}}, \bibinfo{pages}{3030} (\bibinfo{year}{2015}), \eprint{1410.7438}.

\bibitem[{\citenamefont{{Randall} et~al.}(2008)\citenamefont{{Randall}, {Markevitch}, {Clowe}, {Gonzalez}, and {Brada{\v{c}}}}}]{2008ApJ...679.1173R}
\bibinfo{author}{\bibfnamefont{S.~W.} \bibnamefont{{Randall}}}, \bibinfo{author}{\bibfnamefont{M.}~\bibnamefont{{Markevitch}}}, \bibinfo{author}{\bibfnamefont{D.}~\bibnamefont{{Clowe}}}, \bibinfo{author}{\bibfnamefont{A.~H.} \bibnamefont{{Gonzalez}}}, \bibnamefont{and} \bibinfo{author}{\bibfnamefont{M.}~\bibnamefont{{Brada{\v{c}}}}}, \bibinfo{journal}{\apj} \textbf{\bibinfo{volume}{679}}, \bibinfo{pages}{1173} (\bibinfo{year}{2008}), \eprint{0704.0261}.

\bibitem[{\citenamefont{{Gruber} and {Pelster}}(2014)}]{2014EPJD...68..341G}
\bibinfo{author}{\bibfnamefont{C.}~\bibnamefont{{Gruber}}} \bibnamefont{and} \bibinfo{author}{\bibfnamefont{A.}~\bibnamefont{{Pelster}}}, \bibinfo{journal}{European Physical Journal D} \textbf{\bibinfo{volume}{68}}, \bibinfo{eid}{341} (\bibinfo{year}{2014}), \eprint{1403.3812}.

\bibitem[{\citenamefont{{Dalfovo} et~al.}(1999)\citenamefont{{Dalfovo}, {Giorgini}, {Pitaevskii}, and {Stringari}}}]{1999RvMP...71..463D}
\bibinfo{author}{\bibfnamefont{F.}~\bibnamefont{{Dalfovo}}}, \bibinfo{author}{\bibfnamefont{S.}~\bibnamefont{{Giorgini}}}, \bibinfo{author}{\bibfnamefont{L.~P.} \bibnamefont{{Pitaevskii}}}, \bibnamefont{and} \bibinfo{author}{\bibfnamefont{S.}~\bibnamefont{{Stringari}}}, \bibinfo{journal}{Reviews of Modern Physics} \textbf{\bibinfo{volume}{71}}, \bibinfo{pages}{463} (\bibinfo{year}{1999}), \eprint{cond-mat/9806038}.

\bibitem[{\citenamefont{{Chavanis} and {Harko}}(2012)}]{2012PhRvD..86f4011C}
\bibinfo{author}{\bibfnamefont{P.-H.} \bibnamefont{{Chavanis}}} \bibnamefont{and} \bibinfo{author}{\bibfnamefont{T.}~\bibnamefont{{Harko}}}, \bibinfo{journal}{\prd} \textbf{\bibinfo{volume}{86}}, \bibinfo{eid}{064011} (\bibinfo{year}{2012}), \eprint{1108.3986}.

\bibitem[{\citenamefont{{Anderson} et~al.}(1995)\citenamefont{{Anderson}, {Ensher}, {Matthews}, {Wieman}, and {Cornell}}}]{1995Sci...269..198A}
\bibinfo{author}{\bibfnamefont{M.~H.} \bibnamefont{{Anderson}}}, \bibinfo{author}{\bibfnamefont{J.~R.} \bibnamefont{{Ensher}}}, \bibinfo{author}{\bibfnamefont{M.~R.} \bibnamefont{{Matthews}}}, \bibinfo{author}{\bibfnamefont{C.~E.} \bibnamefont{{Wieman}}}, \bibnamefont{and} \bibinfo{author}{\bibfnamefont{E.~A.} \bibnamefont{{Cornell}}}, \bibinfo{journal}{Science} \textbf{\bibinfo{volume}{269}}, \bibinfo{pages}{198} (\bibinfo{year}{1995}).

\bibitem[{\citenamefont{Davis et~al.}(1995{\natexlab{a}})\citenamefont{Davis, Mewes, Andrews, van Druten, Durfee, Kurn, and Ketterle}}]{PhysRevLett.75.3969}
\bibinfo{author}{\bibfnamefont{K.~B.} \bibnamefont{Davis}}, \bibinfo{author}{\bibfnamefont{M.~O.} \bibnamefont{Mewes}}, \bibinfo{author}{\bibfnamefont{M.~R.} \bibnamefont{Andrews}}, \bibinfo{author}{\bibfnamefont{N.~J.} \bibnamefont{van Druten}}, \bibinfo{author}{\bibfnamefont{D.~S.} \bibnamefont{Durfee}}, \bibinfo{author}{\bibfnamefont{D.~M.} \bibnamefont{Kurn}}, \bibnamefont{and} \bibinfo{author}{\bibfnamefont{W.}~\bibnamefont{Ketterle}}, \bibinfo{journal}{Phys. Rev. Lett.} \textbf{\bibinfo{volume}{75}}, \bibinfo{pages}{3969} (\bibinfo{year}{1995}{\natexlab{a}}), \urlprefix\url{https://link.aps.org/doi/10.1103/PhysRevLett.75.3969}.

\bibitem[{\citenamefont{Davis et~al.}(1995{\natexlab{b}})\citenamefont{Davis, Mewes, Joffe, Andrews, and Ketterle}}]{PhysRevLett.74.5202}
\bibinfo{author}{\bibfnamefont{K.~B.} \bibnamefont{Davis}}, \bibinfo{author}{\bibfnamefont{M.-O.} \bibnamefont{Mewes}}, \bibinfo{author}{\bibfnamefont{M.~A.} \bibnamefont{Joffe}}, \bibinfo{author}{\bibfnamefont{M.~R.} \bibnamefont{Andrews}}, \bibnamefont{and} \bibinfo{author}{\bibfnamefont{W.}~\bibnamefont{Ketterle}}, \bibinfo{journal}{Phys. Rev. Lett.} \textbf{\bibinfo{volume}{74}}, \bibinfo{pages}{5202} (\bibinfo{year}{1995}{\natexlab{b}}), \urlprefix\url{https://link.aps.org/doi/10.1103/PhysRevLett.74.5202}.

\bibitem[{\citenamefont{Tiesinga et~al.}(1996)\citenamefont{Tiesinga, Williams, Julienne, Jones, Lett, and Phillips}}]{141041}
\bibinfo{author}{\bibfnamefont{E.}~\bibnamefont{Tiesinga}}, \bibinfo{author}{\bibfnamefont{C.}~\bibnamefont{Williams}}, \bibinfo{author}{\bibfnamefont{P.}~\bibnamefont{Julienne}}, \bibinfo{author}{\bibfnamefont{K.}~\bibnamefont{Jones}}, \bibinfo{author}{\bibfnamefont{P.}~\bibnamefont{Lett}}, \bibnamefont{and} \bibinfo{author}{\bibfnamefont{W.}~\bibnamefont{Phillips}}, \emph{\bibinfo{title}{A spectroscopic determination of scattering lengths for sodium atom collisions}} (\bibinfo{year}{1996}).

\bibitem[{\citenamefont{{Boesten} et~al.}(1997)\citenamefont{{Boesten}, {Tsai}, {Gardner}, {Heinzen}, and {Verhaar}}}]{1997PhRvA..55..636B}
\bibinfo{author}{\bibfnamefont{H.~M.~J.~M.} \bibnamefont{{Boesten}}}, \bibinfo{author}{\bibfnamefont{C.~C.} \bibnamefont{{Tsai}}}, \bibinfo{author}{\bibfnamefont{J.~R.} \bibnamefont{{Gardner}}}, \bibinfo{author}{\bibfnamefont{D.~J.} \bibnamefont{{Heinzen}}}, \bibnamefont{and} \bibinfo{author}{\bibfnamefont{B.~J.} \bibnamefont{{Verhaar}}}, \bibinfo{journal}{\pra} \textbf{\bibinfo{volume}{55}}, \bibinfo{pages}{636} (\bibinfo{year}{1997}).

\bibitem[{\citenamefont{{Dutta} and {Strigari}}(2019)}]{2019ARNPS..69..137D}
\bibinfo{author}{\bibfnamefont{B.}~\bibnamefont{{Dutta}}} \bibnamefont{and} \bibinfo{author}{\bibfnamefont{L.~E.} \bibnamefont{{Strigari}}}, \bibinfo{journal}{Annual Review of Nuclear and Particle Science} \textbf{\bibinfo{volume}{69}}, \bibinfo{pages}{137} (\bibinfo{year}{2019}), \eprint{1901.08876}.

\bibitem[{\citenamefont{{Ruppin} et~al.}(2014)\citenamefont{{Ruppin}, {Billard}, {Figueroa-Feliciano}, and {Strigari}}}]{2014PhRvD..90h3510R}
\bibinfo{author}{\bibfnamefont{F.}~\bibnamefont{{Ruppin}}}, \bibinfo{author}{\bibfnamefont{J.}~\bibnamefont{{Billard}}}, \bibinfo{author}{\bibfnamefont{E.}~\bibnamefont{{Figueroa-Feliciano}}}, \bibnamefont{and} \bibinfo{author}{\bibfnamefont{L.}~\bibnamefont{{Strigari}}}, \bibinfo{journal}{\prd} \textbf{\bibinfo{volume}{90}}, \bibinfo{eid}{083510} (\bibinfo{year}{2014}), \eprint{1408.3581}.

\end{thebibliography}







 
\end{document}